\documentclass[12pt,preprint]{aastex}

\usepackage{color,amsmath,amsfonts,amssymb,amsthm}

\usepackage{mark_newcommands}
\usepackage{aas_macros}  
\usepackage{mark_journals}

\begin{document} 
\title{Baroclinic Vorticity Production in Protoplanetary Disks\\Part II: Vortex Growth and Longevity}
\author{Mark R. Petersen}
\affil{Los Alamos National Laboratory\\  Computer and Computational Science Div. and Center for Nonlinear Studies}
\email{mpetersen@lanl.gov}
\author{Glen R. Stewart}
\affil{Laboratory for Atmospheric and Space Physics, University of Colorado at Boulder}
\author{Keith Julien}
\affil{Dept. of Applied Mathematics, University of Colorado at Boulder}


\begin{abstract}

The factors affecting vortex growth in convectively stable protoplanetary disks are explored using numerical simulations of a two-dimensional anelastic-gas model which includes baroclinic vorticity production and radiative cooling.  The baroclinic feedback, where anomalous temperature gradients produce vorticity through the baroclinic term and vortices then reinforce these temperature gradients, is found to be an important process in the rate of growth of vortices in the disk.  Factors which strengthen the baroclinic feedback include fast radiative cooling, high thermal diffusion, and large radial temperature gradients in the background temperature.  When the baroclinic feedback is sufficiently strong, anticyclonic vortices form from initial random perturbations and maintain their strength for the duration of the simulation, for over 600 orbital periods.  

Based on both simulations and a simple vortex model, we find that the local angular momentum transport due to a single vortex may be inward or outward, depending its orientation.  The global angular momentum transport is highly variable in time, and is sometimes negative and sometimes positive.   This result is for an anelastic gas model, and does not include shocks that could affect angular momentum transport in a compressible-gas disk.


\end{abstract}

\keywords{accretion, accretion disks, circumstellar matter, hydrodynamics, instabilities, methods: numerical, turbulence,  solar system: formation}



\section{Introduction \label{s_introduction}}

The baroclinic term is a source of vorticity in the vorticity equation, and is derived by taking the curl of the pressure gradient in the Navier-Stokes equation,
\bea
\del\times\left(-\frac{1}{\rho} \del p \right)
 = \frac{1}{\rho^2} \del\rho\times\del p,
\label{baroclinic_term}
\eea
where $p$ is the pressure and $\rho$ is the density.  The baroclinic term is nonzero when pressure and density gradients are not aligned.  

An intuitive example of baroclinicity is the land-sea breeze, which is initiated when air temperatures above the land rise more than over the nearby ocean.  The warm air over the land expands, isobars rise relative to those over the ocean, and consequently the isobars tilt towards the ocean.  At the same time, the colder air over the ocean has a higher density than over the land, so the isopycnals tilt towards the land.  The tilting of isobars and isopycnals in opposite directions is a baroclinic source of vorticity, which causes a circulation in the vertical plane that blows from the ocean to the land near the surface.  Thus the potential energy of the tilted isopycnals is converted into kinetic energy of the land-sea breeze, which dissipates through surface friction and reduces the land-sea temperature contrast through temperature advection (see, e.g. \citealp{Holton04bk}).

A related concept is the baroclinic instability, which is of central importance to the production of vortices and Rossby waves at midlatitudes.  Here the decrease in solar insolation from equator to pole causes colder temperatures, and consequently higher density, at the surface at higher latitudes;  thus the isopycnal surfaces are tilted towards the equator.  A system with tilted isopycnals has more potential energy than one with level isopycnals, just like an inclined free surface has more potential energy than a level one.  This potential energy is available to processes which can flatten out the isopycnals.  For example, vortices in the atmosphere and ocean convert the potential energy of the inclined isopycnals to the kinetic energy of their meso-scale motion.  Vortices flatten the isopycnals by transferring heat poleward through their mixing action.  

The baroclinic instability is so-named because of the tilted isopycnals, but the physics is fundamentally different from the land-sea breeze: in the land-sea breeze the circulation is in the vertical plane and caused {\it directly} by the baroclinic term, i.e. by non-aligned density and pressure gradients {\it in the vertical};  in the baroclinic instability the isopycnals are titled in the vertical, but the vortices are in the horizontal plane, so they could not be produced by the baroclinic term directly.  Rather, the tilted isopycnals present an unstable configuration which is ripe for processes which can convert the potential energy to kinetic energy, much like how an avalanche levels out a steep incline of snow.

The baroclinic processes discussed in this paper for a protoplanetary disk are similar to the land-sea breeze, but in radial geometry.  Due to the gravity and radiation of the central star, the density, temperature, and pressure of the disk's gas all decrease radially.  Any azimuthal variations in temperature (and thus density or pressure by the ideal gas law) would lead to an increase in vertical vorticity due to the baroclinic term (\ref{baroclinic_term}).  The focus of this work is the baroclinic feedback, where a vortex enhances azimuthal temperature gradients to reinforce the vortex itself.  Under the right conditions, the baroclinic feedback strengthens vortices so that they can exist for long periods.  These vortices could play a crucial role in planetary formation, as they are efficient at collecting particles from the disk \citep{Tanga_ea96ic, Johansen_ea04aa, Barge_Sommeria95aa, Klahr_Bodenheimer06apj}.  The high density of solids in the vortex would speed the formation by core accretion, which is so slow in the rest of the disk that it may not be a feasible theory of planetary formation there \citep{Wetherill90areps}.  A vortex which collects solids is also a potential site of gravitational instability, where the matter is dense enough that it simply collapses into a planet through gravitation self-attraction \citep{Boss97sc}.

Our study was motivated by \citet{Klahr_Bodenheimer03apj}, who investigated the effects of baroclinicity in a radially stratified disk using a finite difference model of the compressible Navier-Stokes equation combined with a radiative transfer model.   They found the baroclinic instability to be a source of vigorous turbulence which leads to the formation of long-lasting vortices and positive angular momentum transport.  Barotropic simulations, where the entropy (temperature) is constant in the radial direction did not develop turbulence, even with large initial perturbations.  
To explain these results, \citet{Klahr04apj} performed a local linear analysis for a disk with constant surface density, and found that modes do not grow if the growth time of the instability is longer than the shear time.

The issue of whether the baroclinic instability is a mechanism for nonlinear growth and the formation of vortices has been a recent source of debate.  Johnson and Gammie are critical of the findings of \citet{Klahr_Bodenheimer03apj} and \citet{Klahr04apj}.  Their linear analysis found no exponentially growing instabilities, except for convective instabilities in the absence of shear \citep{Johnson_Gammie05apj}.  Furthermore, they use a shearing sheet numerical model to show that disks with a nearly-Keplerian rotation profile and radial gradients on the order of the disk radius are stable to local nonaxisymmetric disturbances \citep{Johnson_Gammie06apj}.  

The goal of this study is to understand the effects of baroclinic instabilities and radiative cooling on the generation of turbulence, vortex formation, and vortex longevity in protoplanetary disks.  One of our motivations is to shed light on the conflicting observations of baroclinic instabilities by \citet{Klahr_Bodenheimer03apj} and \citet{Johnson_Gammie06apj}.  

This work is presented in two parts.  Part I, which precedes this article, presents the equation set, details of the numerical model, and results of the small-domain simulations, which are used to study the process of vortex formation.  This paper, Part II, explores the parameters which affect the baroclinic feedback during the growth phase of the vortices; these simulations use the larger quarter annulus domain and are run for hundreds of orbital periods to observe the long-term behavior of the vortices.  We begin with a quick review of the equation set in \S \ref{s_cuppd_equations}.  The results in \S \ref{s_results} discuss the   evolution of a typical simulation, the process of the baroclinic feedback, the Richardson number as a diagnostic, and the alpha viscosity.  In \S \ref{s_angular_momentum_transport} we discuss the angular momentum transport in our simulations, which is highly variable and depends on the orientation of individual vortices.  In \S \ref{s_conclusions} we conclude that the baroclinic instability is an important mechanism for vortex generation and persistence, and review the conditions which affect the instability.  For conciseness there is little repetition between Parts I and II, so the reader is advised to read both together.

\section{Description of the Equation Set \label{s_cuppd_equations}}
The model equations are described fully in Part I of this work, and are only briefly reviewed here.  They model an anelastic gas, which filter out pressure waves that restrict the timestep of the numerical model but do not impact the physics of interest here.  Our equations set is similar to those in \citet{Bannon96jas} and \citet{Scinocca_Shepherd92jas}, which are anelastic models of the atmosphere derived from conservation of momentum, conservation of mass, the second law of thermodynamics, and the ideal gas law.  Our equations use two-dimensional polar coordinates $(r,\phi)$ where temperature and density are stratified in the radial direction.  Variables such as the vertical component of vorticity $\zeta$, streamfunction $\Psi$, potential temperature $\theta$, thermal temperature $T$, surface density $\Sigma$, and Exner pressure $\pi$ are written as the sum of a background and perturbation term, e.g. $\theta=\theta_0(r)+\theta'(r,\phi,t)$, where the background functions only vary radially.

The model equations are
\bea
\zeta' &=& \ds \frac{1}{r} \frac{\p}{\p r}
   \left( \frac{r}{\Sigma_0} \frac{\p \Psi'}{\p r} \right)
   + \frac{1}{r^2 \Sigma_0} \frac{\p^2\Psi'}{\p \phi^2} 
\label{num_stream_fxn}\\
\ds \frac{\p\zeta'}{\p t}
  + \p\left(\Psi,\frac{1}{\Sigma_0}\zeta\right) \ds
  &=& \ds  \frac{c_p}{r}
      \dd{\pi_0}{r}\dd{\theta'}{\phi} 
  + \nu_e \del^2 \zeta' 
\label{num_vorticity} \\
\ds \frac{\p \theta'}{\p t}
  + \frac{1}{\Sigma_0}\p\left(\Psi,\theta\right)
  &=& \ds -\frac{\theta'}{\tau} + \kappa_e \del^2 \theta'.
\label{num_temperature}
\eea
The first is the relationship between the perturbation streamfunction $\Phi'$ and the perturbation vorticity $\zeta'$; the second and third are prognostic equations for perturbation vorticity $\zeta'$ and perturbation potential temperature $\theta'$.  Radial and azimuthal velocities $\bu=(u,v)$ are related to the streamfunction by $\Sigma_0 \bu = -\del\times\Psi {\bf \hat{z}}$.  Other variables include the radiative cooling time $\tau$, specific heat at constant pressure $c_p$, time $t$, viscosity $\nu_e$, thermal dissipation $\kappa_e$, vertical unit vector ${\bf \hat{z}}$, and the Jacobian $\p(a,b)=(\p_ra\p_\phi b - \p_\phi a\p_rb)/r$.

The baroclinic term, 
\bea
\frac{c_p}{r} \dd{\pi_0}{r}\dd{\theta'}{\phi},
\eea
is a central focus of this paper.  It is the only source term in the vorticity equation (\ref{num_vorticity}), and it plays an important role in the baroclinic instability, as one might expect.  Most people are familiar with the baroclinic term using density and pressure, shown in eqn. (\ref{baroclinic_term}).  This operation in terms of our variables is
\bea
\del\times\left(-c_p \theta'\del\pi_0 \right)
 =-\frac{1}{r}\frac{\p}{\p\phi}\left(-c_p \theta'\frac{d\pi_0}{dr} \right)
 = \frac{c_p}{r}\frac{\p\theta'}{\p\phi} \frac{d\pi_0}{dr}.
\eea
The factor $\p_r{\pi_0}$ indicates that the baroclinic feedback should be strengthened if $\p_r{\pi_0}$ is large, i.e., if radial pressure gradients are large.  But
\bea
\dd{p_0}{r} \sim \dd{\Sigma_0 T_0}{r}
= \Sigma_0 \dd{T_0}{r} + T_0 \dd{\Sigma_0}{r},
\label{pres_temp_density}
\eea
so we expect large radial temperature or density gradients to strengthen the baroclinic feedback.

The radiative cooling term $\theta'/\tau$ diffuses the perturbation potential temperature equally at all scales with an e-folding time of $\tau$.  The two Laplacian terms, $\nu_e\del^2\zeta'$ and $\kappa_e\del^2\theta'$, diffuse potential vorticity and potential temperature at fastest at the highest wavenumbers.  They were added to the numerical model to dissipate energy for numerical stability.  In the nondimensionalized version of the equation set the $\nu$ coefficients are replaced with the Reynolds and Peclet numbers,
\beas
Re = \frac{L_{sc}^2}{\nu_e t_{sc}}, \hspace{1cm}
Pe = \frac{L_{sc}^2}{\kappa_e t_{sc}},
\eeas
where the length scale $L_{sc}$ and time scale $t_{sc}$ are described in Part I.

\section{Results\label{s_results}}

The simulations discussed in this paper vary parameters such as the radiative cooling rate, the background temperature and surface density gradients, and the Peclet number (Table \ref{t_parameters}).  These simulations capture the salient features of the physics of the anelastic equation set.  The topic of Part I was vortex formation, and thus used a smaller domain for only five orbital periods.  Here we are interested in vortex growth and longevity due to the baroclinic feedback, and have chosen a larger domain and durations of 300 to 600 orbital periods. (This is 6,200 to 12,400 years for a solar-mass star.)  The simulations were performed on the quarter annulus with a radial extent from 5AU to 10AU and a resolution of $256\times256$ and $512\times512$ gridpoints.

The background surface density and background temperature are constant in time and are power functions in the radial,
\bea
\Sigma_0(r) = a\left(\frac{r}{r_{in}}\right)^b, \;\;\; 
T_0(r) = c\left(\frac{r}{r_{in}}\right)^d,
\eea
where $r_{in}=5AU$ is the inner radius of the annulus.  The coefficients are $a=1000$g cm$^{-2}$, $c=600$K for the quarter annulus domain; $b$ and $d$ are varied and shown in Table \ref{t_parameters}.  For example, for simulation A1, the background surface density varies from 1000g cm$^{-2}$ to 350g cm$^{-2}$ and the background temperature decreases radially from 600K to 150K.  This range of temperatures can only be achieved in a realistic disk when the radius ranges from 1 AU to 10 AU \citep{Boss98areps}.  We have artificially enhanced the radial temperature gradient in order to compensate for the lower resolution of our global simulations.  We have demonstrated in paper I using a higher-resolution local simulation that more realistic temperature gradients can still produce vortices.  Most simulations were run to 300 orbital periods, measured as a full ($2\pi$) orbit at $r_{mid}=7.5$AU.  This is 6,200 years for a solar-mass star.

Thermal temperature $T$ and potential temperature $\theta$ are related by
\bea
\theta=T\left(\frac{p_0(r_{in})}{p}\right)^{R/c_p}=\frac{T}{\pi},
\eea
where $R$ is the gas constant and $\pi$ is the Exner pressure.  All results in this paper are expressed in terms of the thermal temperature $T$ in order to compare to observations.  The potential temperature is a measure of entropy.  If entropy increases radially ($d\theta_0/dr>0$) then the disk is convectively stable---this is the Schwarzschild criterion \citep{Schwarzschild58bk}.  If the entropy gradient as accompanied by differential rotation, the Solberg-H{\o}iland criterion \citep{Tassoul00bk,Ruediger_ea02aa} is used to test convective stability (see Results section of Part I).  For the simulations presented in this paper, the Solberg-H{\o}iland value is positive (0.035--0.299 years$^{-2}$), indicating that they are convectively stable.

\begin{figure}[tbh]
\center
\scalebox{.15}{\includegraphics{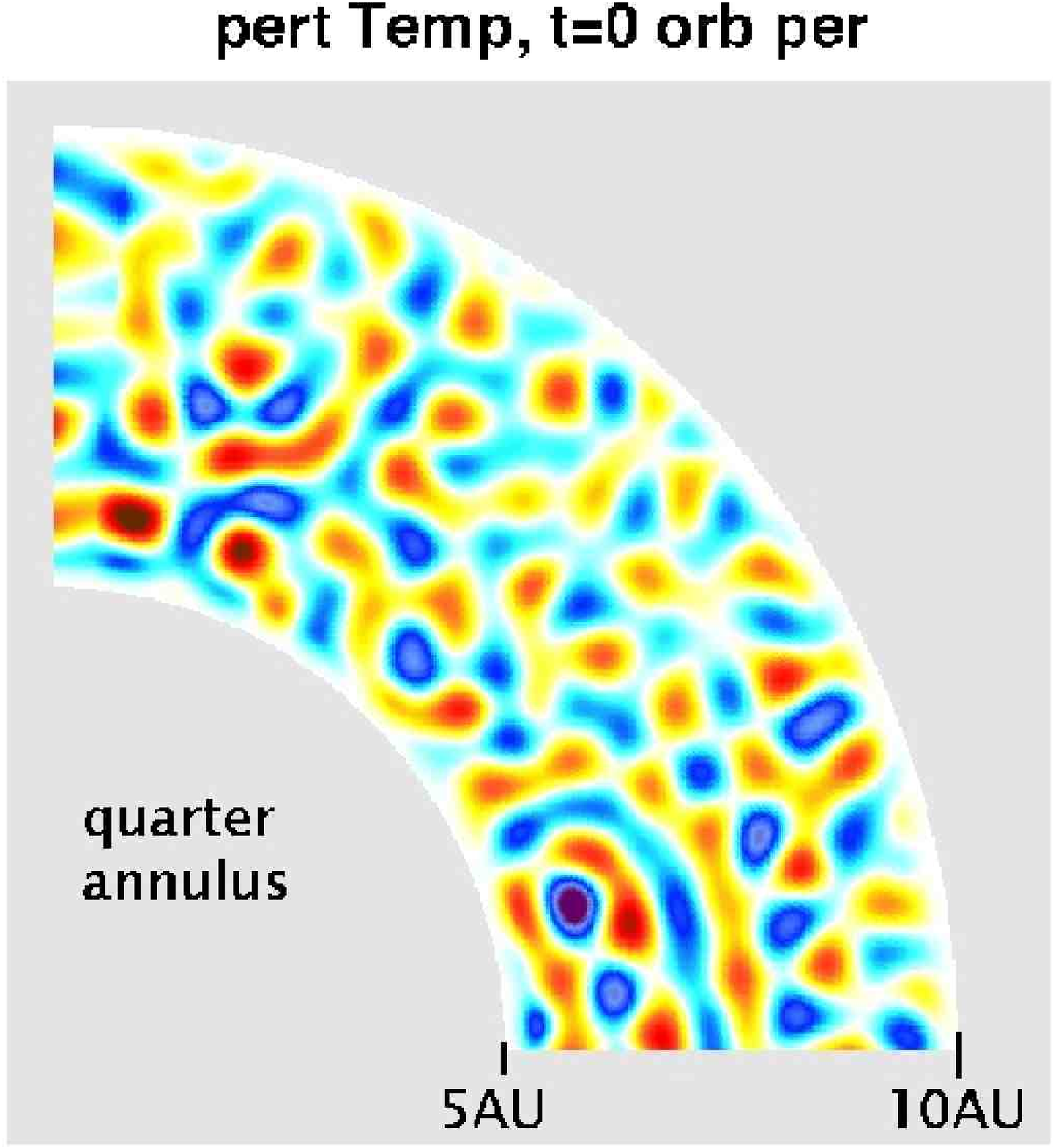}}  
\scalebox{.4}{\includegraphics{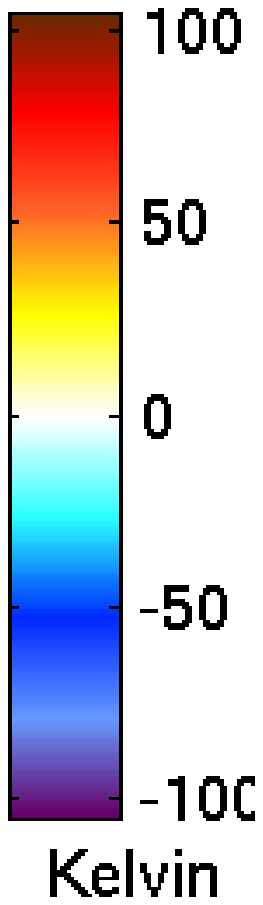}}  
\caption{\label{f_ic} 
Initial temperature perturbation $T'$.}
\end{figure}

The initial condition for the perturbation temperature is shown in Fig. \ref{f_ic}.  It is created with a specified wavenumber distribution in Fourier space, transformed to Cartesian coordinates, and interpolated to the Fourier-Chebyshev annular grid (see Part I, section 3.2).  The initial vorticity perturbation is created in a similar fashion.  The magnitude of the initial conditions is 25\% of the maximum of the background function. 

The small domain simulations in Part I ($r\in [9.5,10], \phi\in[0,\pi/32]$) required a much smaller initial perturbation to initiate vortices---a temperature perturbation of only 5\%, and an initial vorticity perturbation of zero.  This is possible because the small domain is a higher resolution relative to the background shear.  The sensitivity analysis in Part I showed that smaller initial perturbations are required to initiate vortices with progressively higher resolution and Reynolds number.  The same is true of the background temperature; the quarter annulus domain uses higher temperatures ($c=600K$) and steeper gradients ($d=-2$) than the small domain simulations.  Again, the sensitivity analysis in Part I showed that at higher resolutions vortices can be formed with progressively cooler disks and shallower background gradients.

\begin{figure}[tbh]
 \center
 \scalebox{.12}{\includegraphics{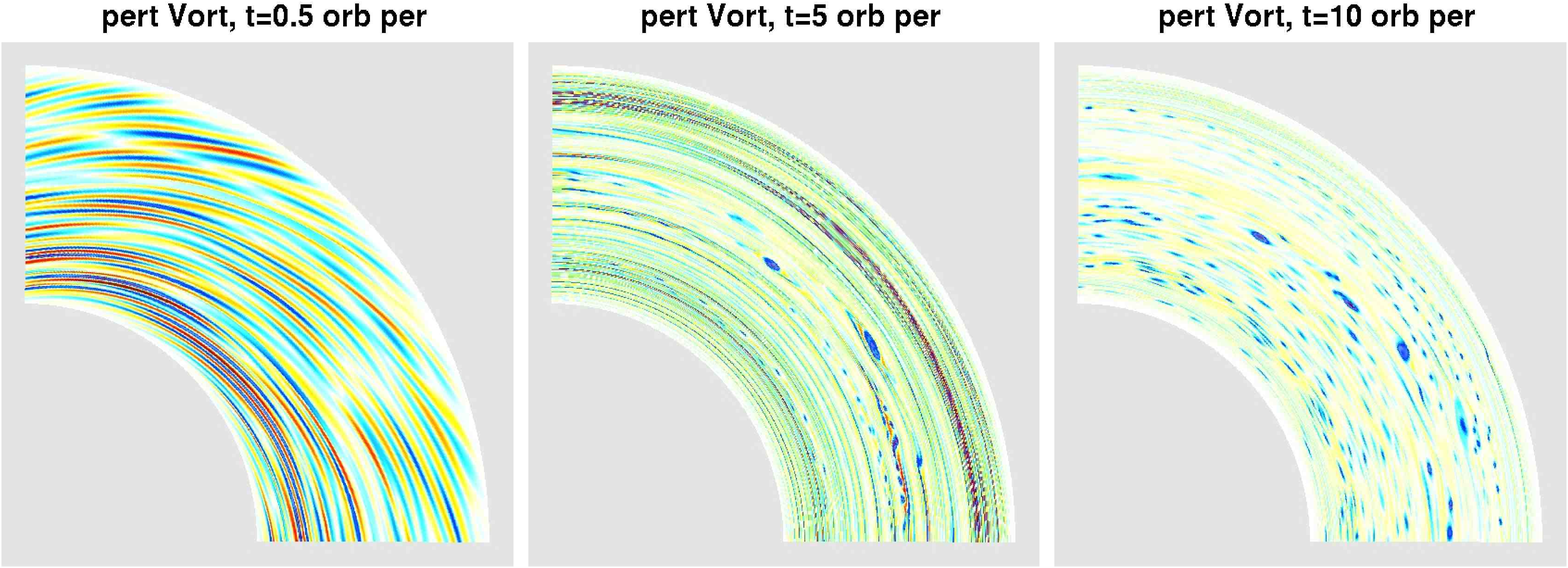}}  
 \scalebox{.4}{\includegraphics{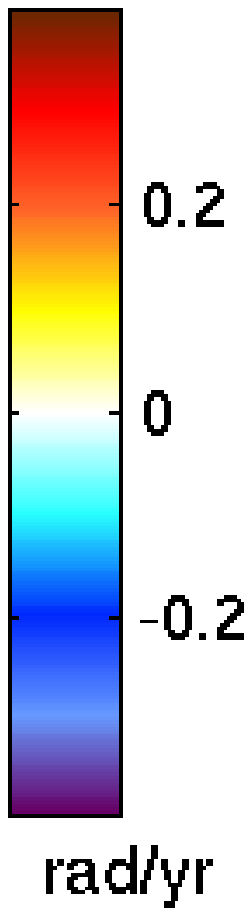}}  
 \caption{\label{f_t10} 
 The perturbation vorticity $\zeta'$ in the quarter-annular computational domain for simulation A1.  The time $t$ refers to the orbital period in the middle of the annulus.  At very early time (left), the vorticity is simply sheared by the background differential rotation.  By five orbital periods (center) a few anti-cyclonic vortices begin to fold over, and by ten orbital periods (right) numerous small anticyclonic vortices have formed.  }
\end{figure}

\begin{figure}[tbh]
\center
\scalebox{.8}{\includegraphics{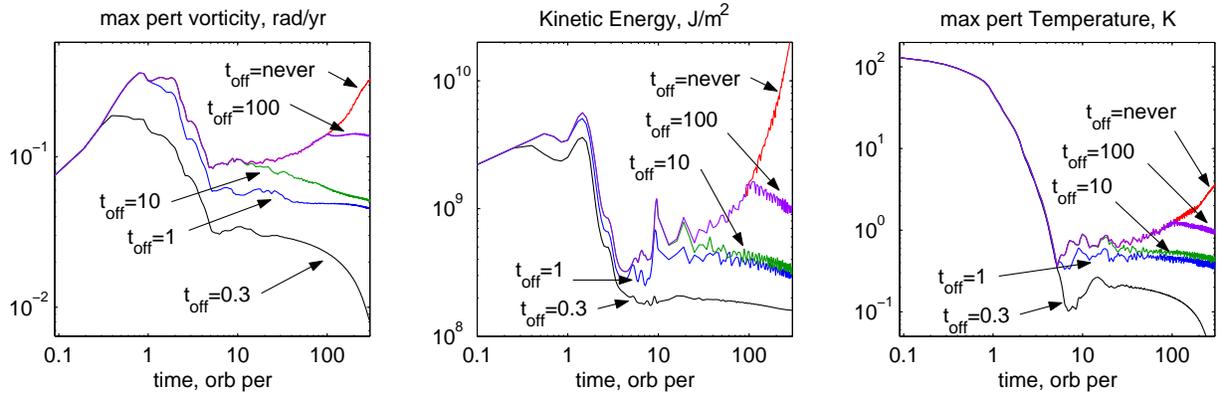}}  
\caption{\label{f_bcl_off} 
Comparison of maximum perturbation vorticity $|\zeta'|$ (left), perturbation kinetic energy (center), and maximum perturbation temperature $|T'|$ (right) for simulation B1, where the baroclinic term is turned off at the times indicated.  When this occurs the vorticity and kinetic energy immediately drop off, indicating that vortex growth is due to the baroclinic term.
}\end{figure}

The evolution of a typical simulation can be described as follows: The initial distribution of vorticity shears due to the differential rotation of the nearly-Keplerian rotational profile (Fig. \ref{f_t10}).  Even at these early times, the perturbation vorticity and perturbation kinetic energy grow due to the baroclinic term (Fig \ref{f_bcl_off}).  After about five orbital periods, anticyclonic vortices begin to form, and by ten orbital periods the domain is populated by numerous small anti-cyclones.  Cyclonic (anti-cyclonic) fluid rotates in the same (opposite) direction as the background fluid, and is denoted by positive (negative) vorticity perturbation in the figures.  It is well-known that anti-cyclones can be long-lived in a Keplerian disk, while cyclones shear out into thin filaments that eventually dissipate away \citep{Godon_Livio99apj, Marcus90jfm,Marcus_ea00pp}.  An anticyclonic vortex has a positive azimuthal velocity at small inner radii and a negative azimuthal velocity at large outer radii.  This means that anticyclonic vortices can smoothly match the background shear flow, and therefore extract energy from the Keplerian shear.  Cyclonic vortices cannot smoothly match the background shear flow and are therefore sheared apart.

\begin{figure}[tbh]
\center
\scalebox{.12}{\includegraphics{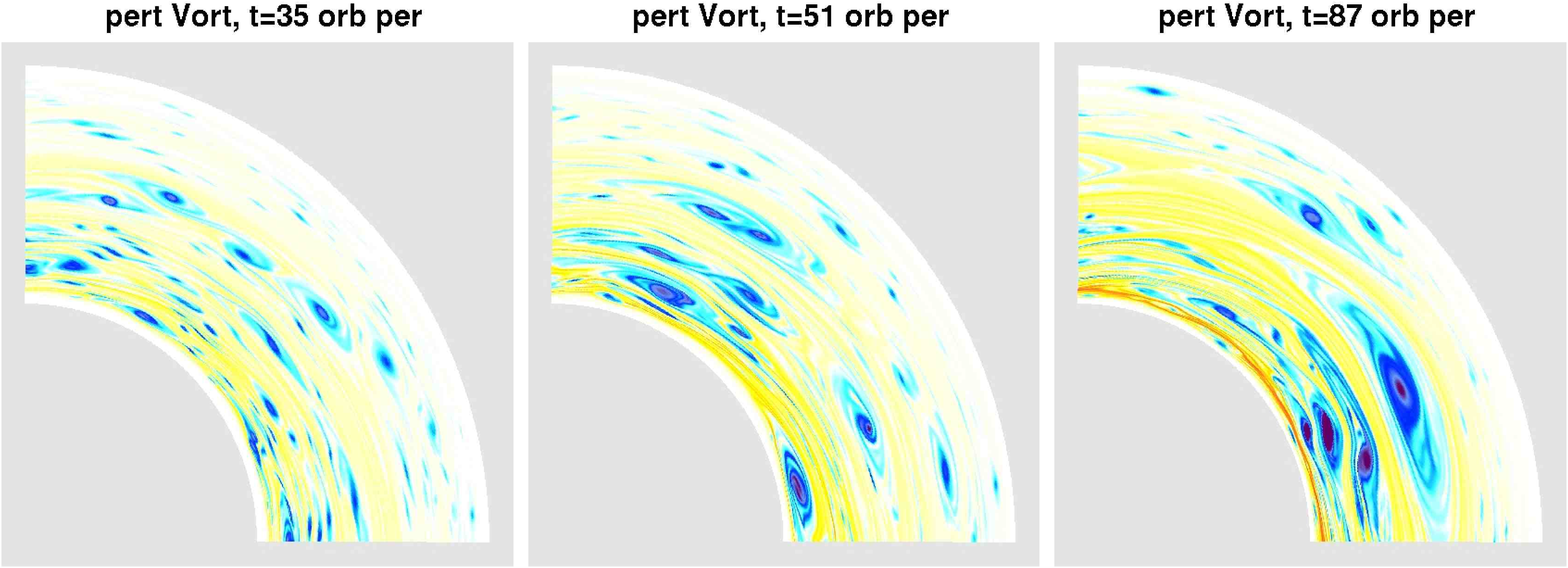}}  
\scalebox{.4}{\includegraphics{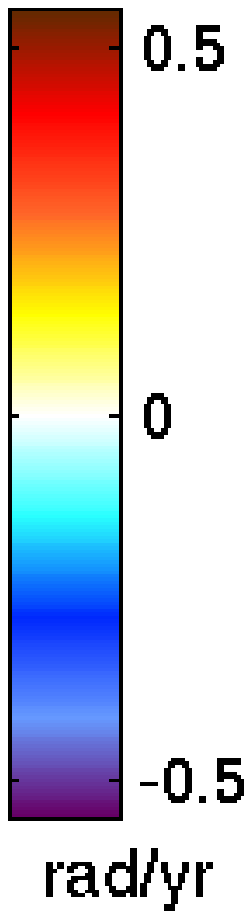}}  
\scalebox{.12}{\includegraphics{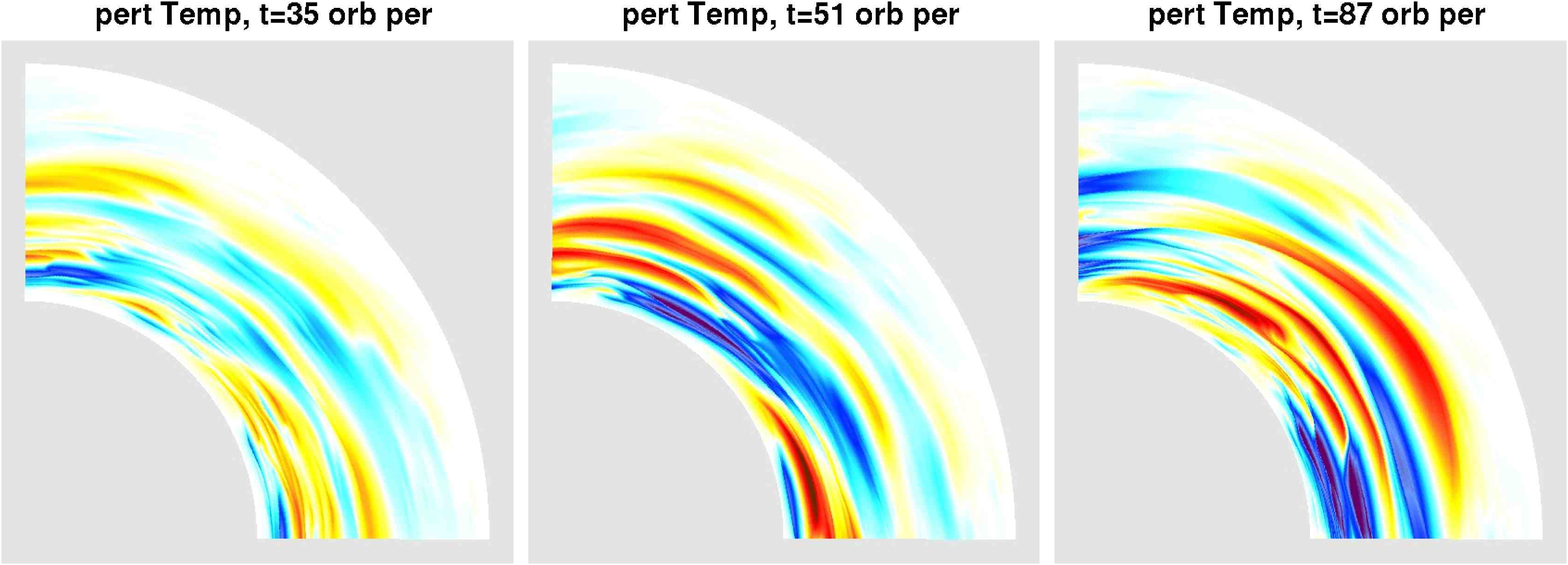}}  
\scalebox{.4}{\includegraphics{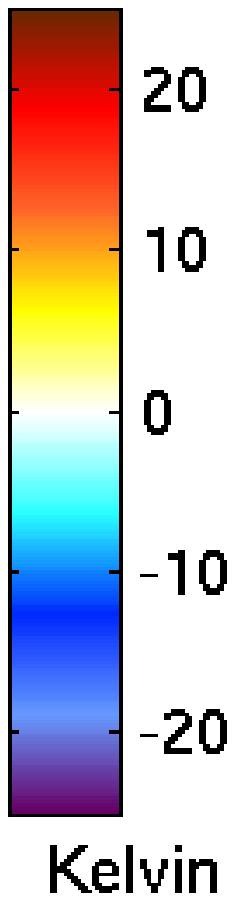}}  
\caption{\label{f_t100}
The perturbation vorticity $\zeta'$ (top) and perturbation temperature $T'$ (bottom) for simulation A1, where the radiative cooling time is fast ($\tau=1$). In this regime the baroclinic feedback remains strong, and vortices remain strong for the full simulation. }
\end{figure}

\begin{figure}[tbh]
\center
\scalebox{.12}{\includegraphics{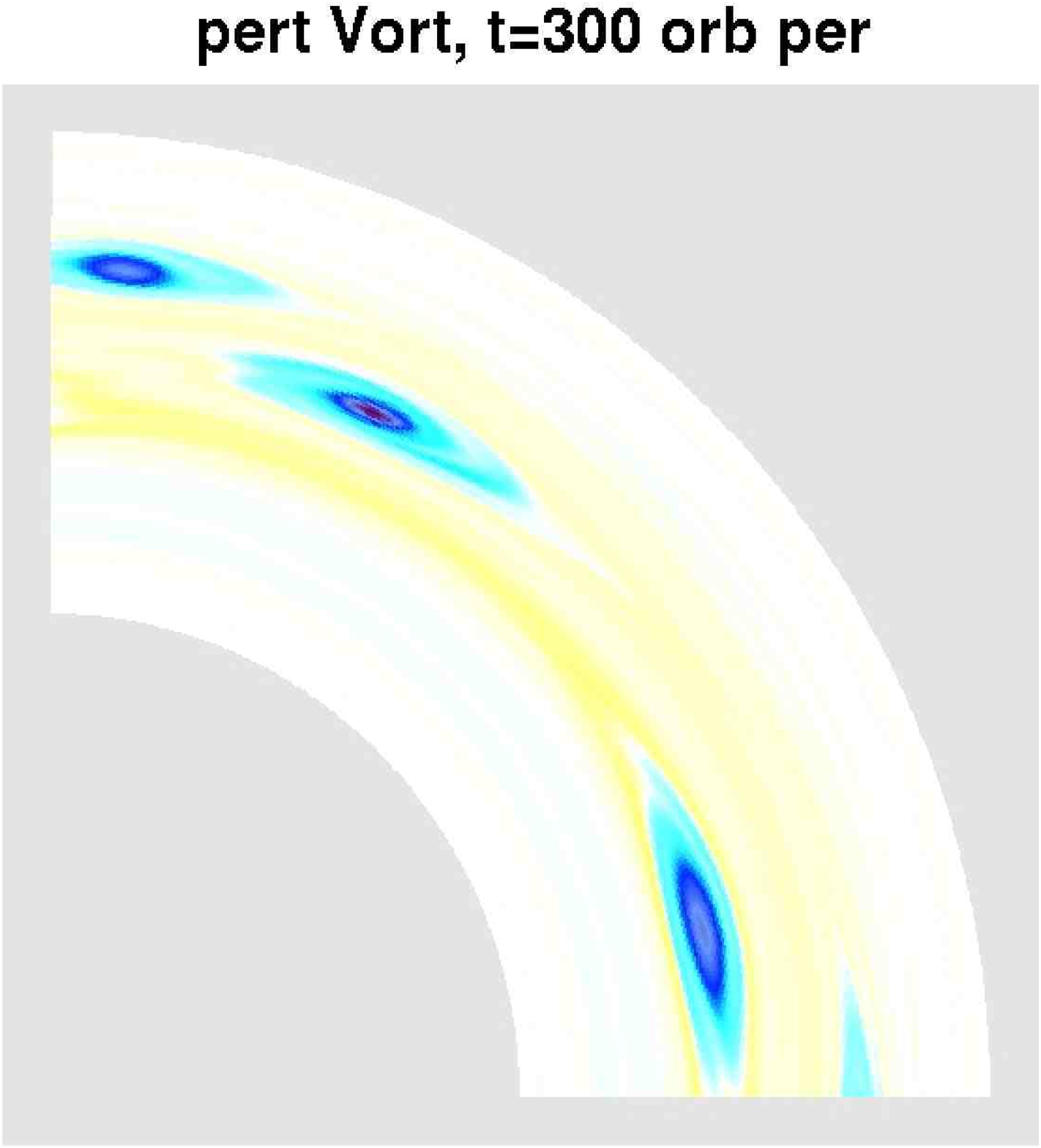}}  
\scalebox{.4}{\includegraphics{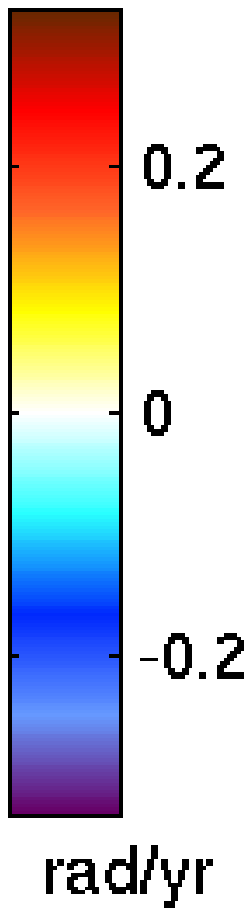}}  
\scalebox{.12}{\includegraphics{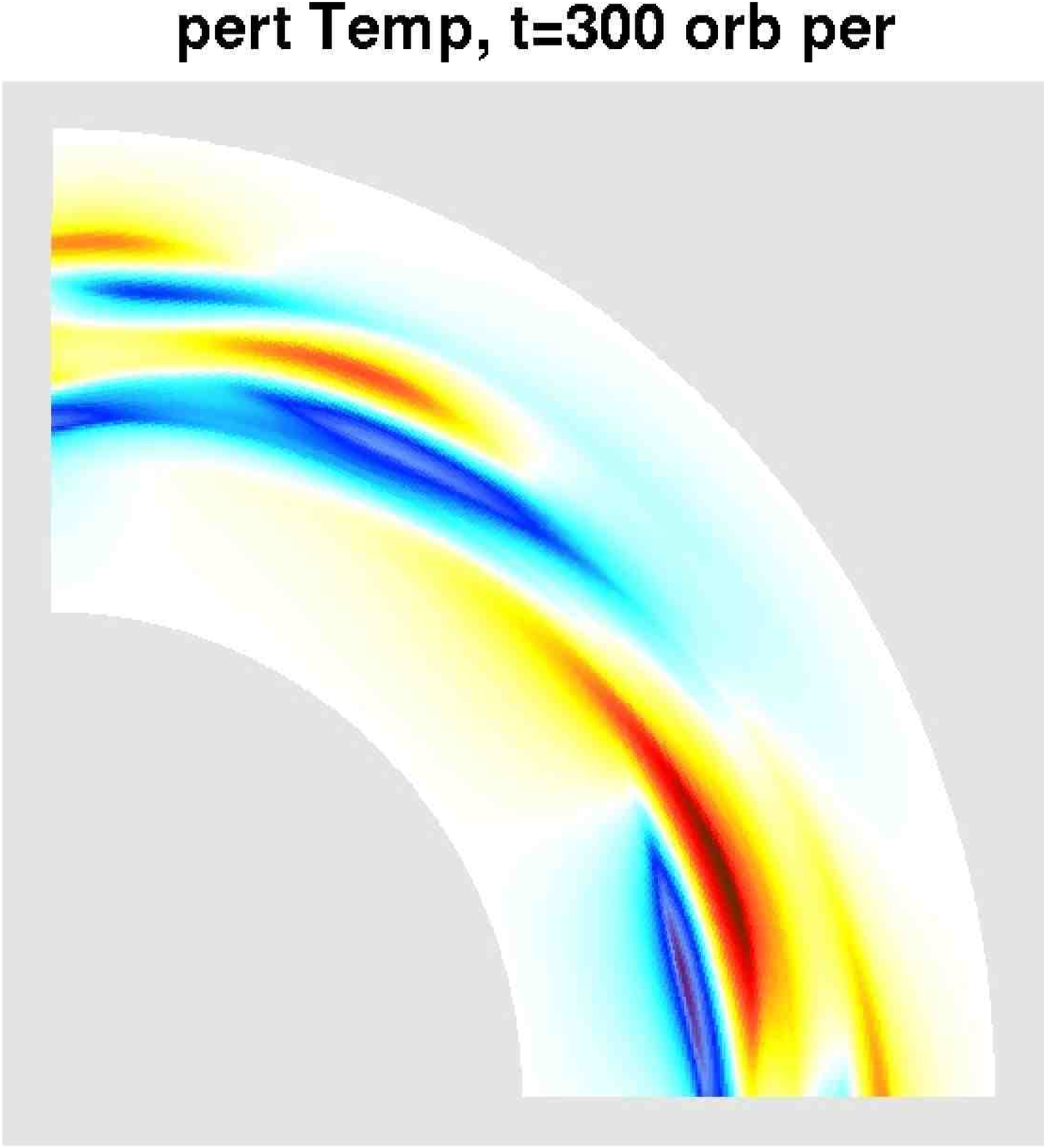}}  
\scalebox{.4}{\includegraphics{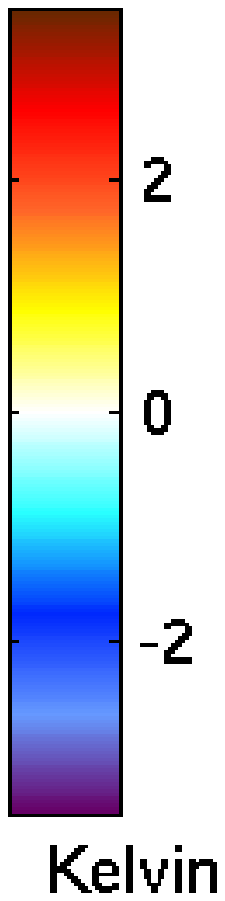}}  
\caption{\label{f_t300} 
Perturbation vorticity and temperature for simulation T3, where $d=-0.75$.  Despite dissipation of vorticity from the numerical code, the vortices remain long-lived because baroclinic vorticity production reinforces the vortices and balances the dissipation.  Here the ``sandwich'' pattern about each vortex is clearly seen: temperature perturbations track each vortex with a warm band to the outside and a cool band to the inside.
} \end{figure}

After the initial period of vortex formation, the vortices merge and grow in strength (Figs. \ref{f_t100}, \ref{f_t300}).  This merging behavior is similar to the merging of like-signed vortices in two-dimensional isotropic turbulence, which transfers energy from smaller to larger scales (the inverse cascade).  However, in shearing flows vortices do not merge as readily and must be sufficiently close in the radial direction.  It is not at all clear 
that this merging of vortices can occur in a fully three-dimensional 
disk if the initial radial vortex scale is small compared to the disk 
scale height.  On the other hand, if vortices primarily form on the 
upper and lower surfaces of a vertical stratified disk as found by 
\citet{Barranco_Marcus05apj}, then it may be possible for small-scale 
vortices to merge in these surface layers.   Further discussion and images of vortex merger, longevity, and distribution can be found in \citet{Godon_Livio99apj} and \citet{Umurhan_Regev04aa}.  

There is a clear ``sandwich'' pattern of temperature perturbations around each vortex (Fig. \ref{f_t300}): the vortex advects warmer fluid towards the outside of the disk and cooler fluid towards the inside of the disk.  In the sandwich analogy, the temperature perturbations are the bread and the vortex is the meat between the bread.  These perturbations have azimuthal temperature gradients that play a role in the baroclinic feedback.

\subsection{Baroclinic vorticity production \label{s_baroclinic_q}}
The model equations for vorticity and temperature perturbations are coupled by the baroclinic term in the vorticity equation (\ref{num_vorticity}) and the advection term in the temperature equation (\ref{num_temperature}).  This coupling is required to support long-lived vortices; without it, vorticity and temperature perturbations simply decay to zero. 

The baroclinic feedback operates as follows: 
\begin{enumerate}
\item Azimuthal gradients in the perturbation temperature field, $\partial\theta'/\partial\phi$, make the baroclinic term in the vorticity equation non-zero.
\item The baroclinic term is a source of vorticity which strengthens anticyclonic vortices.
\item Vortices stir the fluid, moving warm fluid from the inner annulus outward and cool fluid from the outer annulus inward.
\item This local advective heat transport enhances azimuthal temperature gradients, $\partial\theta'/\partial\phi$, completing the feedback cycle.
\end{enumerate}

In order to show that the vortex growth is indeed due to this baroclinic feedback, the baroclinic term was turned off at various times in simulation set B (Fig. \ref{f_bcl_off}).  In all of these trials, perturbation vorticity and kinetic energy drop off immediately when the baroclinic term is turned off.  This is particularly striking at $t=10$  and $t=100$, when vortex strength is growing quickly in the reference simulation.  The kinetic energy in these plots is computed from the perturbation velocity fields.

\begin{figure}[tbh]
\center
\scalebox{.8}{\includegraphics{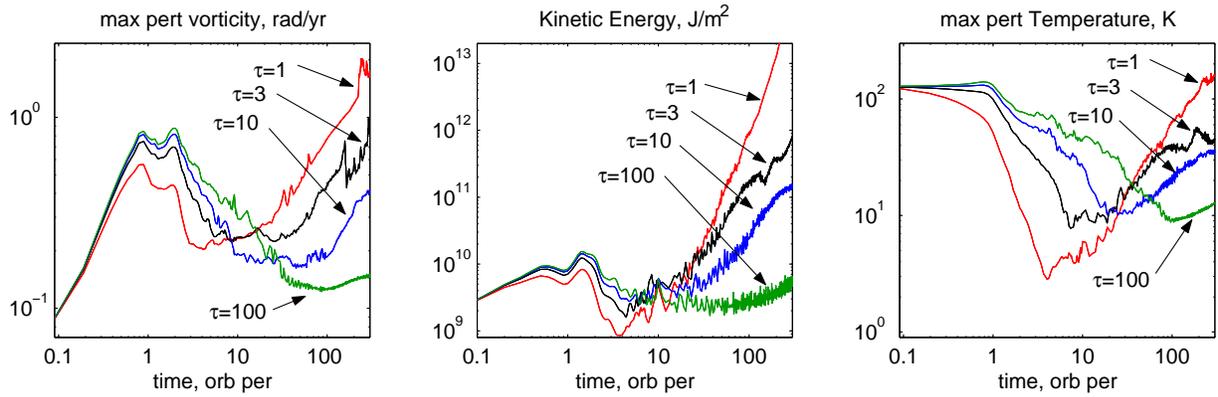}}  
\caption{\label{f_vary_tau} 
Data from simulations Tau1 through Tau4, where $\tau$, the radiative cooling time, varies between 1 and 100.  Two distinct stages can be seen: during vortex formation---at early times---the disk cools rapidly, and cools fastest with small $\tau$; once vortices have formed the baroclinic feedback takes effect, and vortices grow fastest with small $\tau$.
}\end{figure}

The rate of thermal dissipation, $\tau$,  plays a crucial role in the formation and growth of vortices.  Fig. \ref{f_vary_tau} shows that there are two distinct stages in these simulations: vortex formation, from $t=0$ to about $t=5$ orbital periods, and vortex growth, which occurs after $t=5$.  During vortex formation, small thermal dissipation (large $\tau$) allows the strongest vortices to form.  That is because the initial temperature perturbation dies off quickly when thermal dissipation is large, so that azimuthal temperature gradients gradients are smaller and the baroclinic term produces less vorticity.  This is not yet the baroclinic feedback because steps 3 and 4 are missing---it is just baroclinic vorticity production from the initial temperature gradients, which are steps 1 and 2.

\begin{figure}[tbh]
\center
\scalebox{.12}{\includegraphics{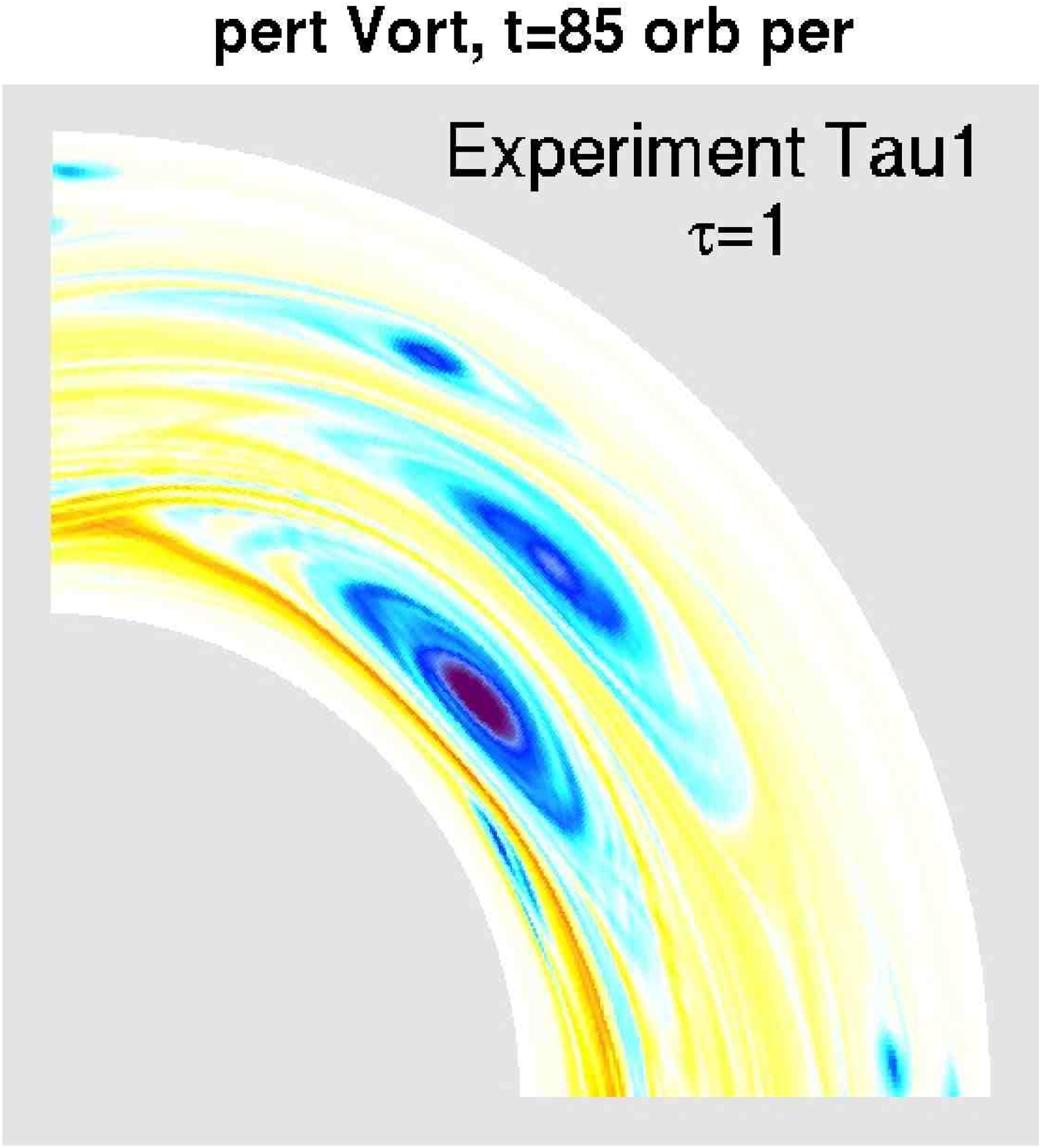}}  
\scalebox{.12}{\includegraphics{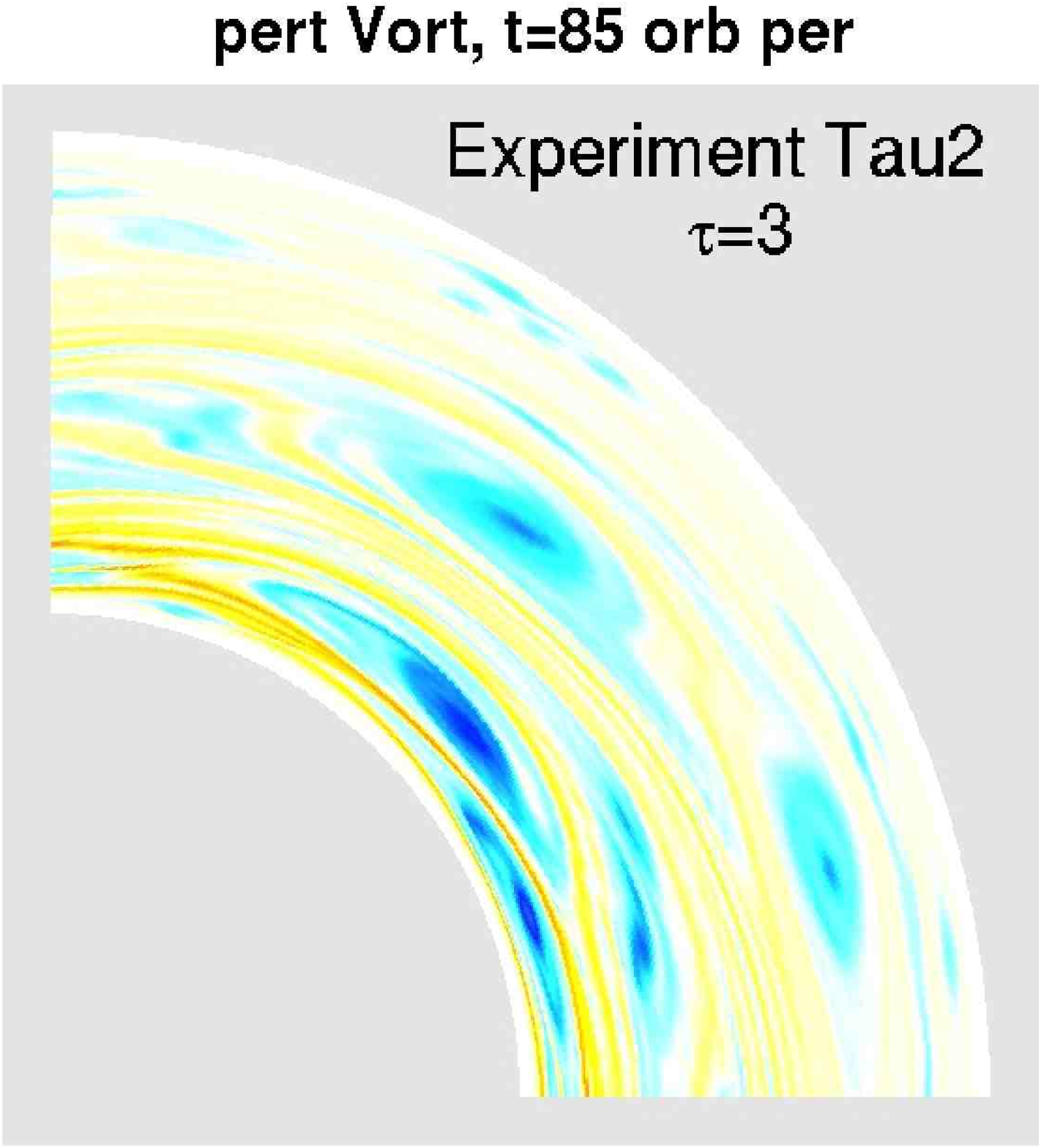}}  
\scalebox{.12}{\includegraphics{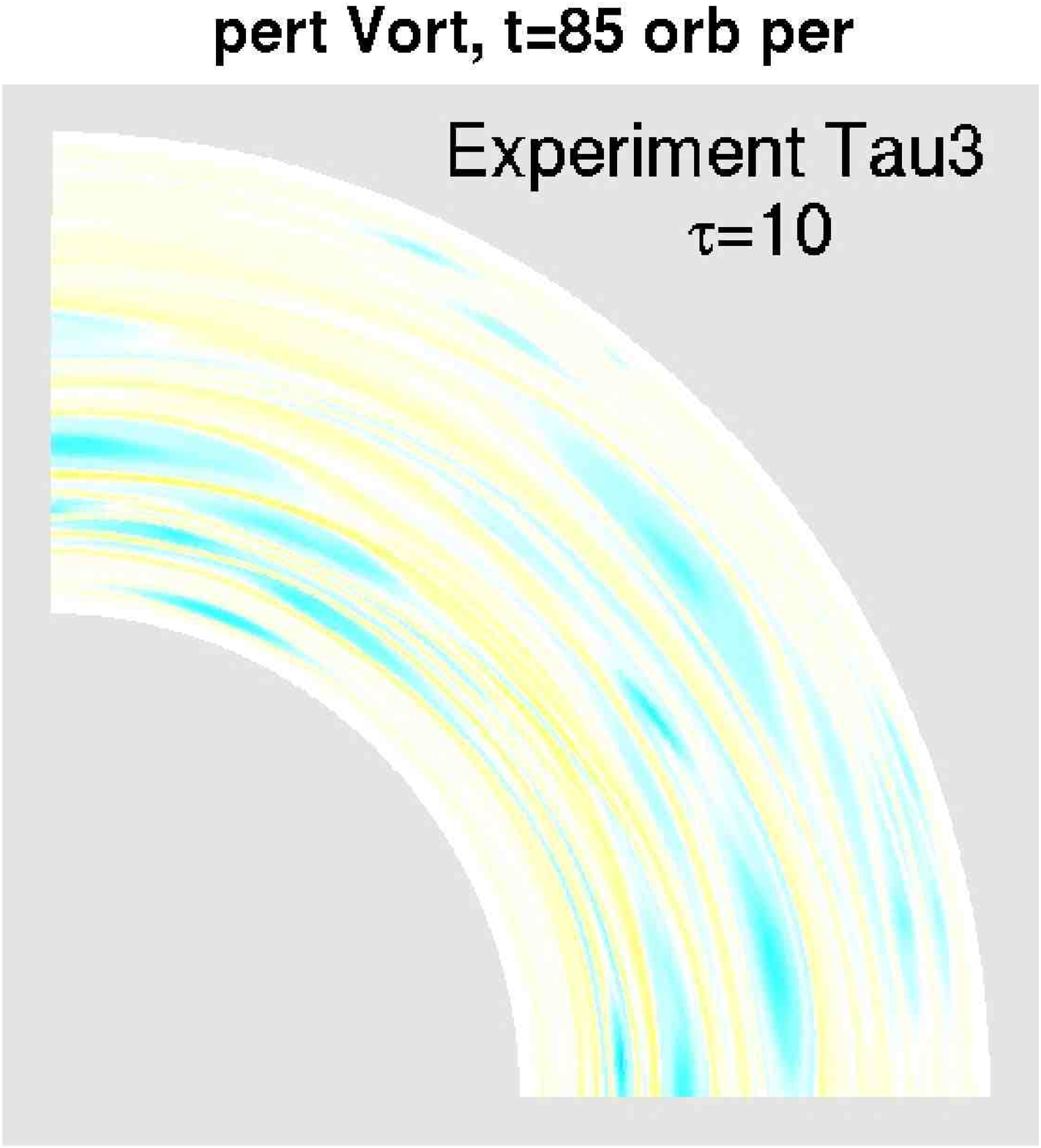}}  
\scalebox{.35}{\includegraphics{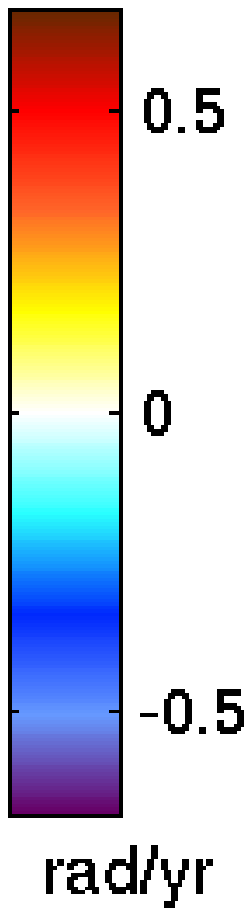}}\\  
\scalebox{.12}{\includegraphics{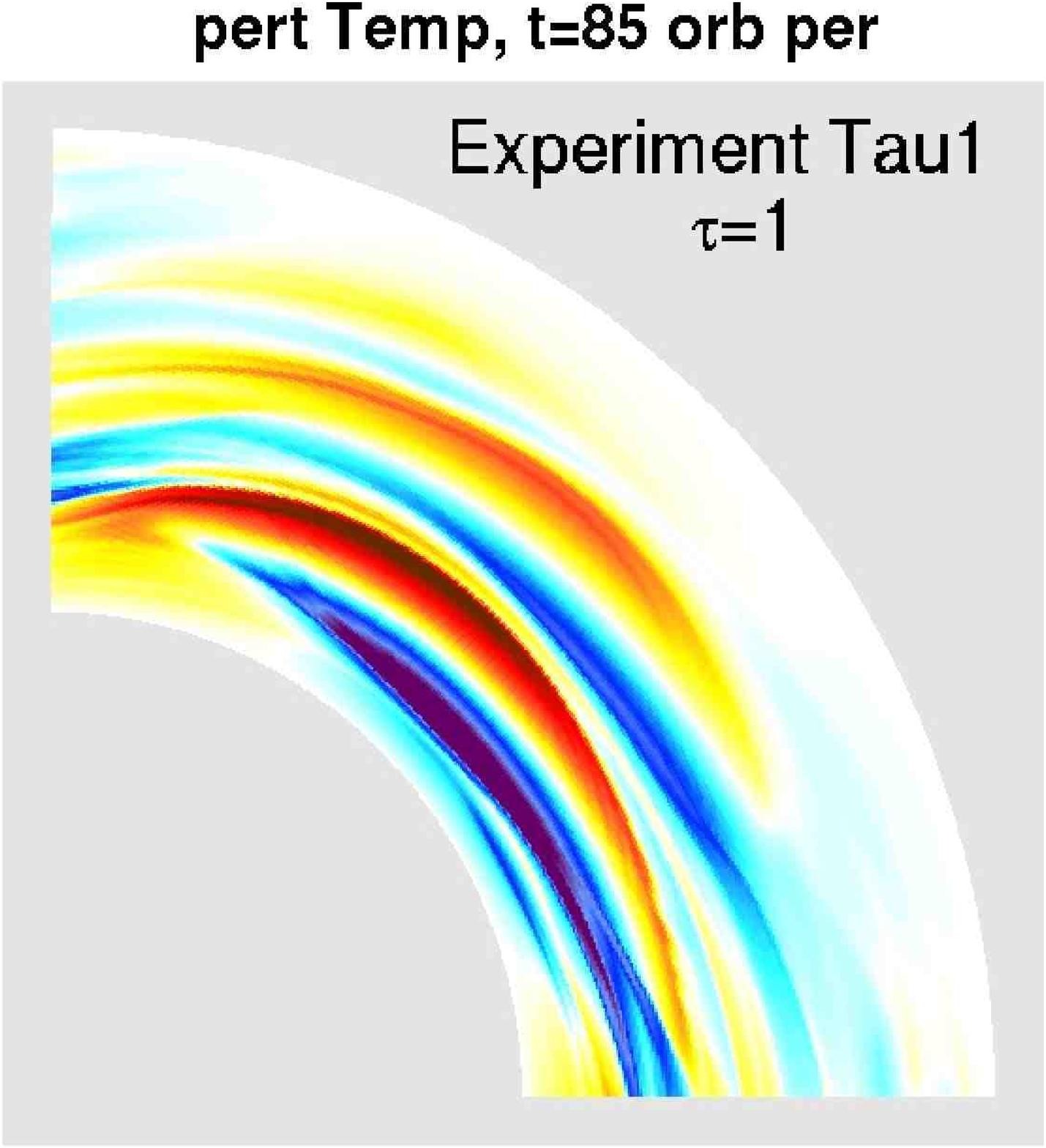}}  
\scalebox{.12}{\includegraphics{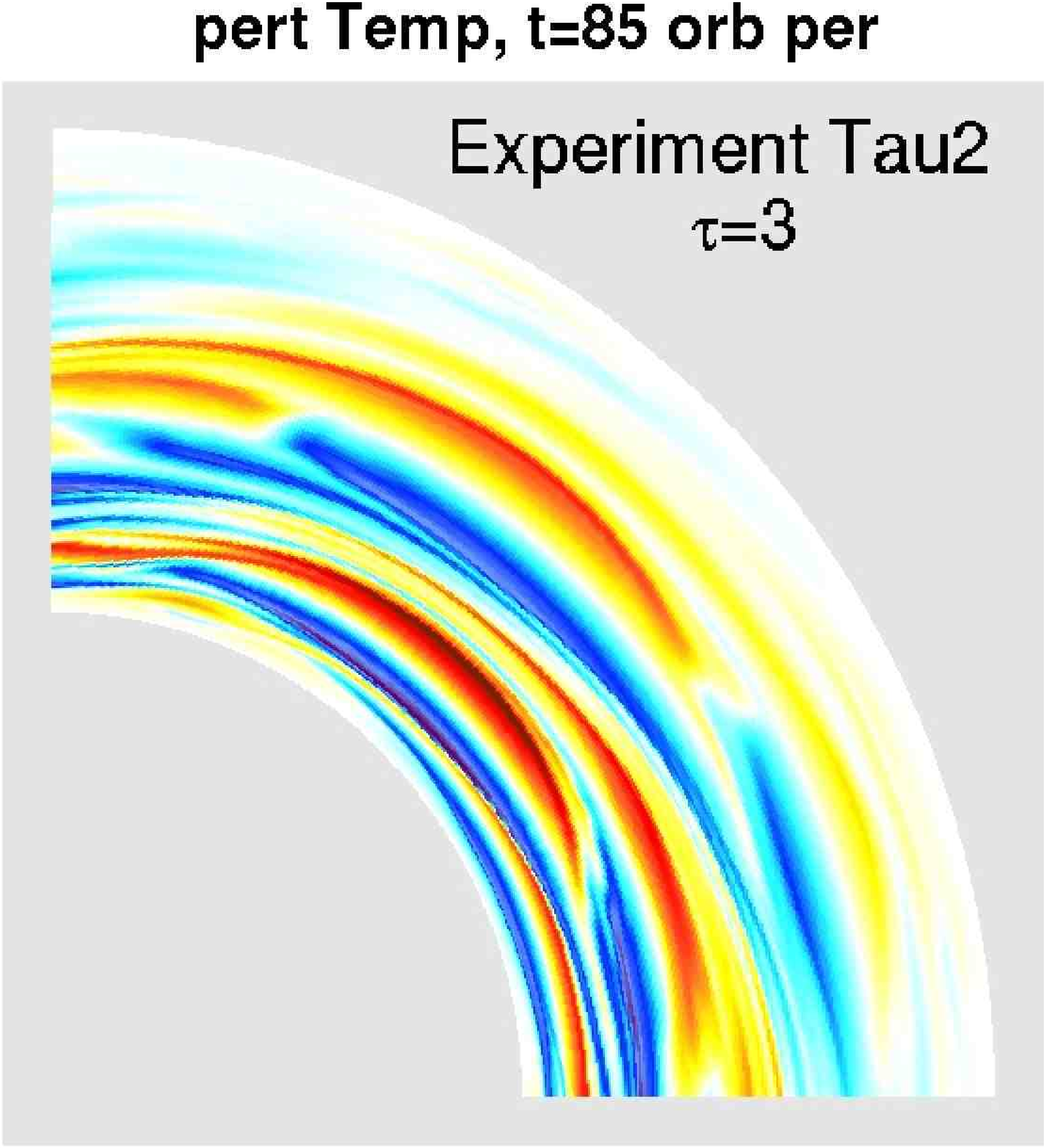}}  
\scalebox{.12}{\includegraphics{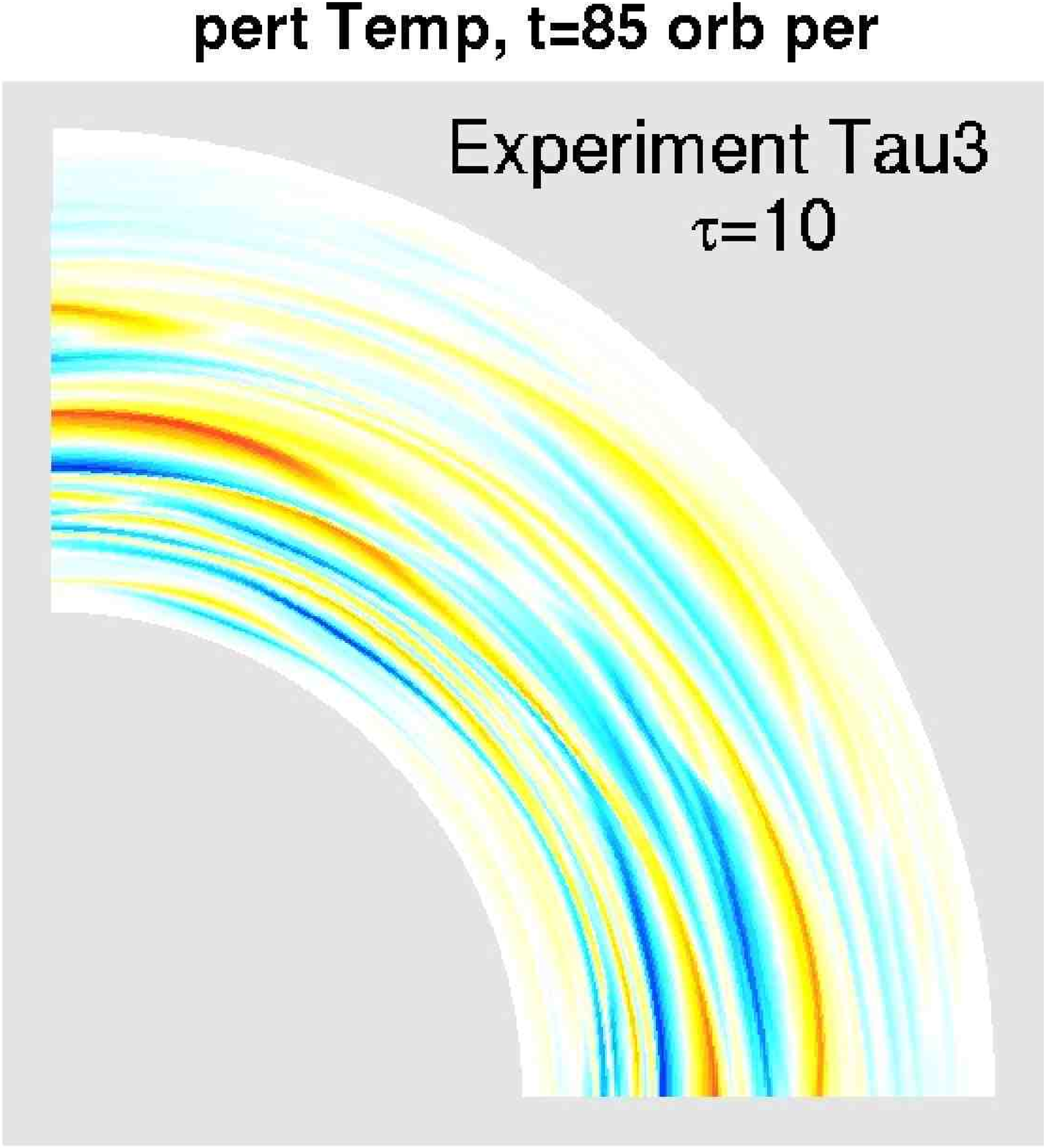}}  
\scalebox{.35}{\includegraphics{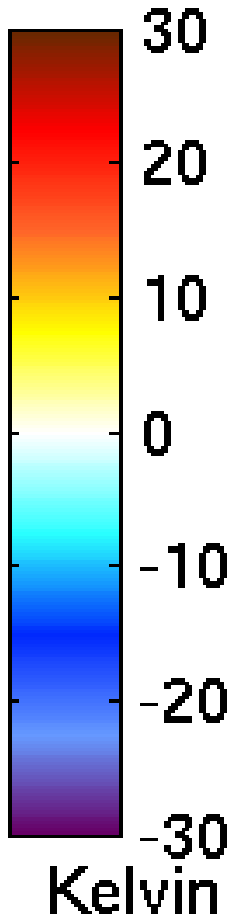}}  
\caption{\label{f_tau} 
Perturbation vorticity and temperature at 85 orbital periods from simulations Tau1, Tau2, and Tau3, where the radiative cooling time $\tau$ is varied from 1 to 10.  When the radiative cooling time is slow (large $\tau$, right), the temperature responds sluggishly to vortices, making the baroclinic feedback weak.  
}\end{figure}

Once vortices form they advect fluid about them (step 3), creating the distinctive ``sandwich'' pattern of cool (warm) temperature perturbations to the inside (outside) of the vortex, as shown in Fig. \ref{f_t300}.  These temperature perturbations create local azimuthal temperature gradients (step 4), completing the cycle of the baroclinic feedback.  Sometime after five orbital periods, the vortices have formed and the simulation transitions from the vortex formation stage to the vortex growth stage.  Now that the baroclinic feedback is operating, thermal dissipation has the opposite effect than at early times (Fig. \ref{f_vary_tau}).
If the disk cools quickly ($\tau$ small), then the warm and cool temperature perturbations can remain tight about each vortex, so that $\partial\theta'/\partial\phi$ in the baroclinic term is large, and the baroclinic feedback is strong.
If the disk cools slowly ($\tau$ large), the perturbation temperature responds sluggishly to mixing by vortices, $\partial\theta'/\partial\phi$ is small, the baroclinic feedback is weak, and vortices simply dissipate away (Fig. \ref{f_tau}).   Quantitative measures of disk activity like kinetic energy, and maximum temperature and vorticity clearly show that the strength of the feedback and rate of growth of vortices is strongly dependent on $\tau$ (Fig. \ref{f_vary_tau}).  In simulations where the radiative cooling rate was sufficiently fast, the baroclinic feedback counters dissipation and vortices remain strong and coherent for hundreds of orbital periods (Fig. \ref{f_t300}).  The longest running simulation, T3 where $d=-0.75$, ended at 600 orbital periods, at which point all of the vortices had merged into a single anticyclonic vortex.  

\begin{figure}[tbh]
\center
\scalebox{.8}{\includegraphics{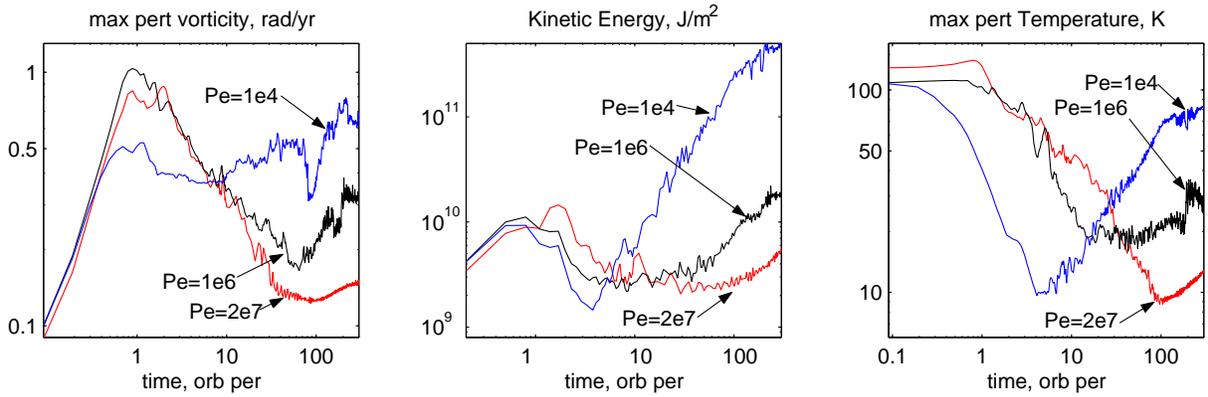}}  
\caption{\label{f_vary_Pe} Data from simulations Pe1--Pe3, which compare the effects of varying Peclet number $Pe$.  High Peclet number indicates low thermal diffusion.  Increasing thermal diffusion (decreasing $Pe$) strengthens the baroclinic feedback, as exemplified by the slopes of the kinetic energy after $t=10$.  This is similar to increasing the radiative cooling rate.
}\end{figure}

\begin{figure}[h]
\center
\scalebox{.12}{\includegraphics{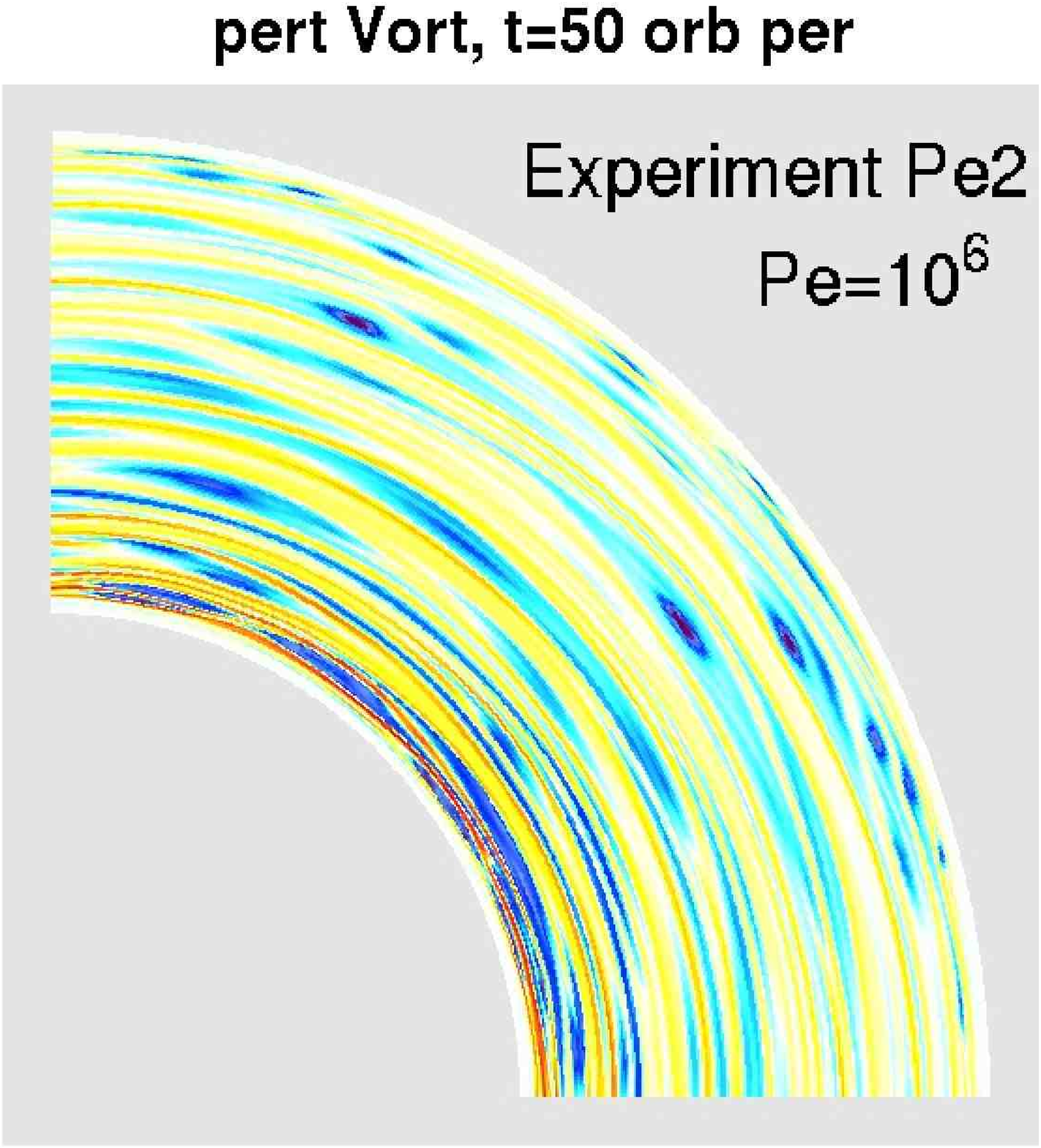}}  
\scalebox{.4}{\includegraphics{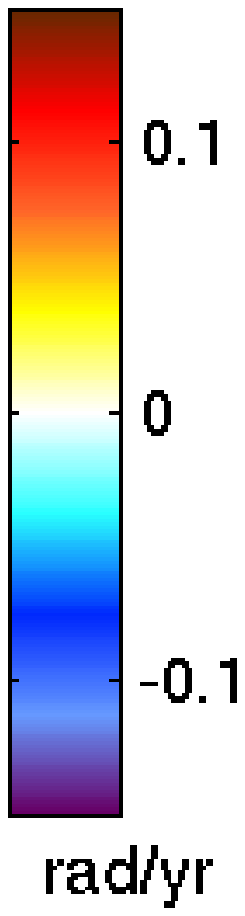}}  
\scalebox{.12}{\includegraphics{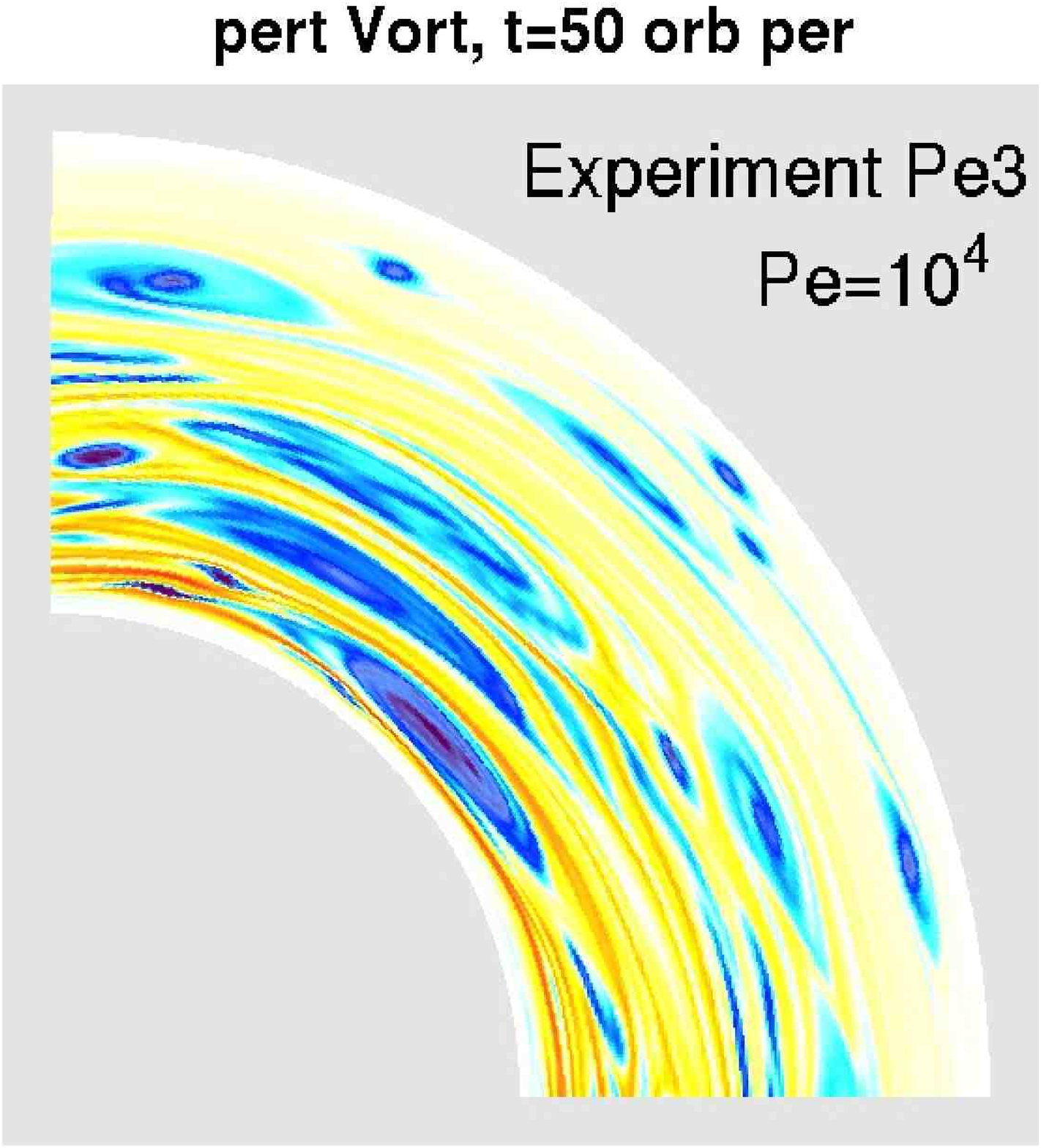}}  
\scalebox{.4}{\includegraphics{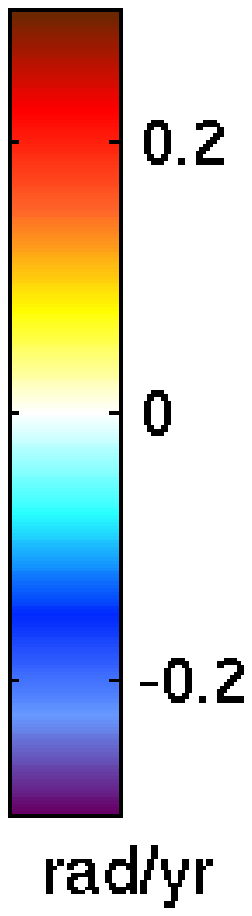}}\\  
\scalebox{.12}{\includegraphics{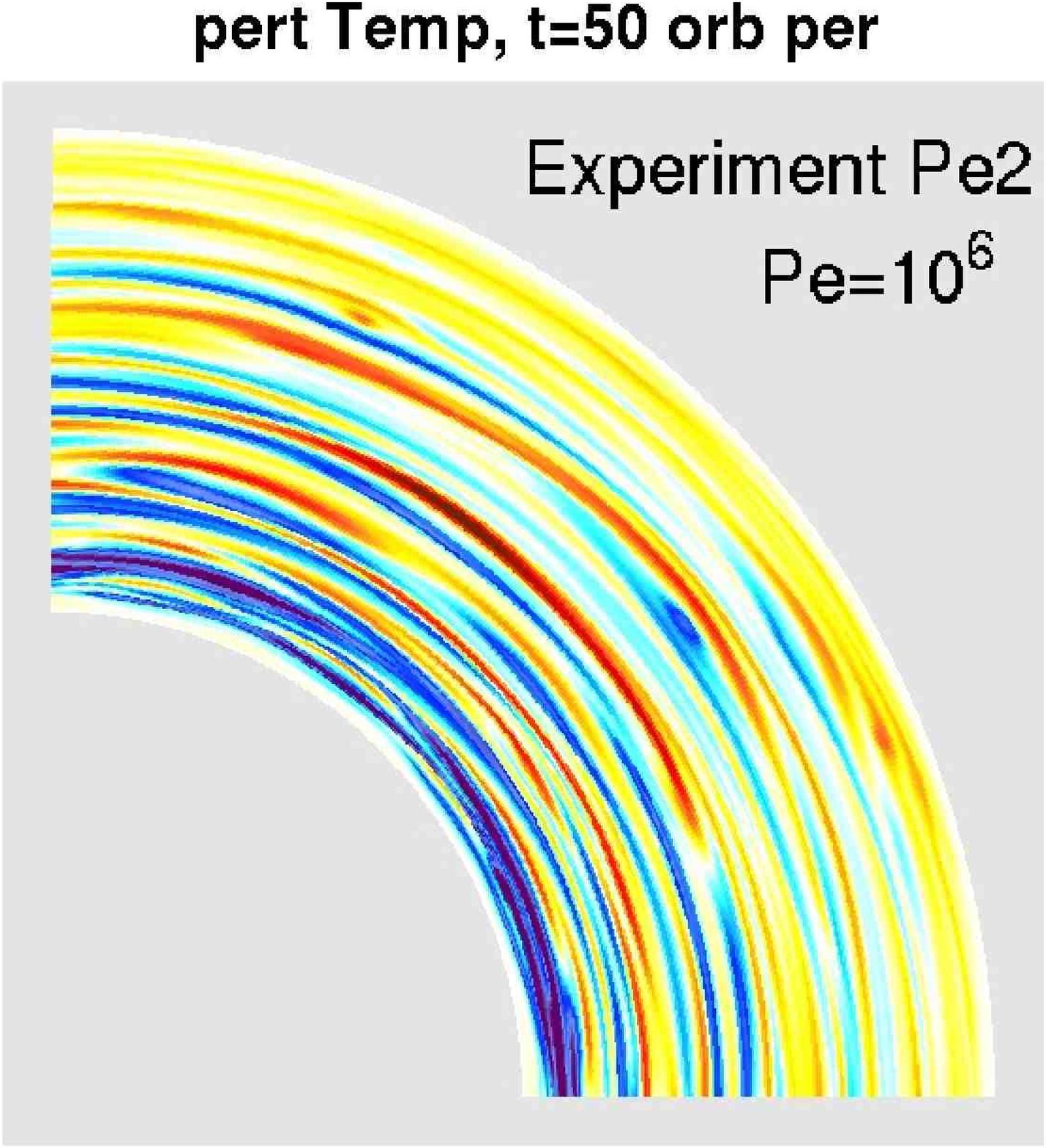}}  
\scalebox{.4}{\includegraphics{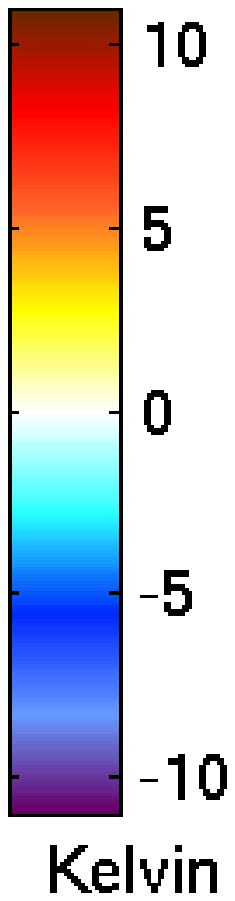}}  
\scalebox{.12}{\includegraphics{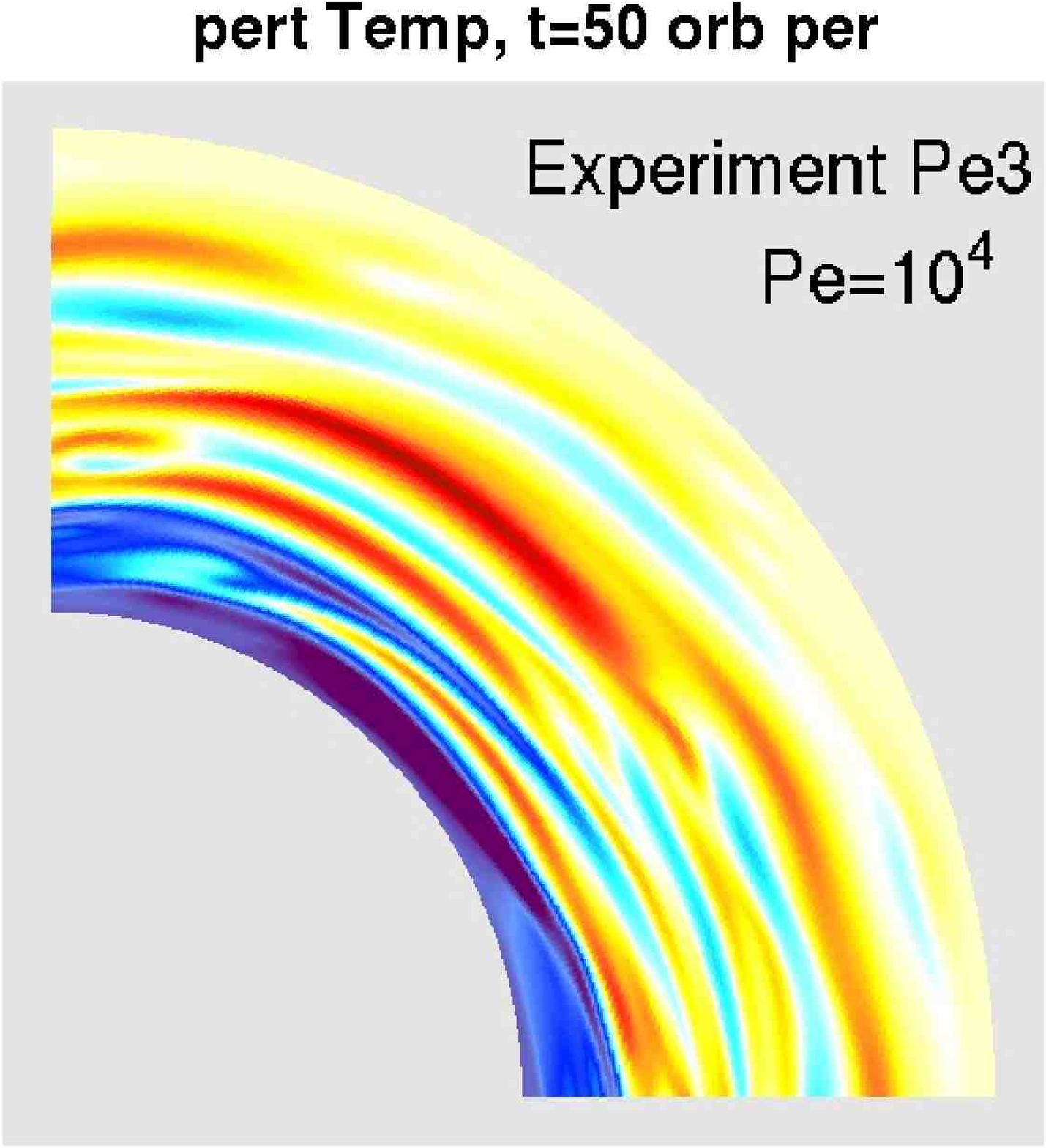}}  
\scalebox{.4}{\includegraphics{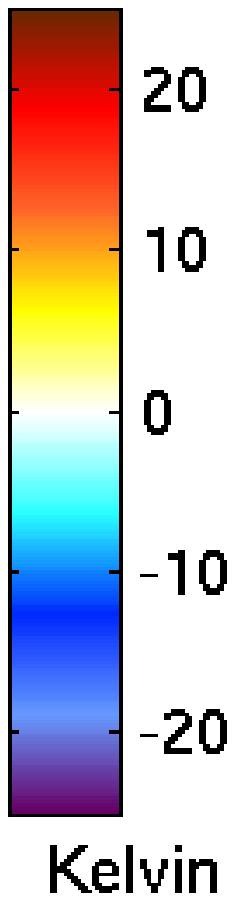}}  
\caption{\label{f_Pe_sections} 
Perturbation vorticity and temperature after 50 orbital periods for simulations Pe2 and Pe3.  Simulation Pe3 (right) has higher thermal diffusion (lower $Pe$), resulting in temperature perturbations that are larger in size (bottom right), and a stronger baroclinic feedback.
}\end{figure}

There are two dissipative terms in the temperature equation (\ref{num_temperature}): the Laplacian term $\kappa_e \nabla^2\theta'$, which dissipates most quickly at small scales; and the radiative cooling term $-\theta'/\tau$ which dissipates equally at all scales.  Can the Laplacian term play the same role as the radiative cooling term in the baroclinic feedback?  Simulations Pe1-Pe3, where $Pe$ ranges from $10^4$ to $2\times10^7$, show that the Laplacian term can indeed play that role (Fig. \ref{f_vary_Pe}); higher thermal diffusion (smaller Peclet number) produces a stronger baroclinic feedback.   Higher diffusion produces warm and cool areas around each vortex which are more localized azimuthally, and therefore have larger azimuthal temperature gradients (Fig. \ref{f_Pe_sections}).  The azimuthal temperature gradients then produce more vorticity through the baroclinic term (step 1 of the baroclinic feedback).

\begin{figure}[h]
\center
\scalebox{.1}{\includegraphics{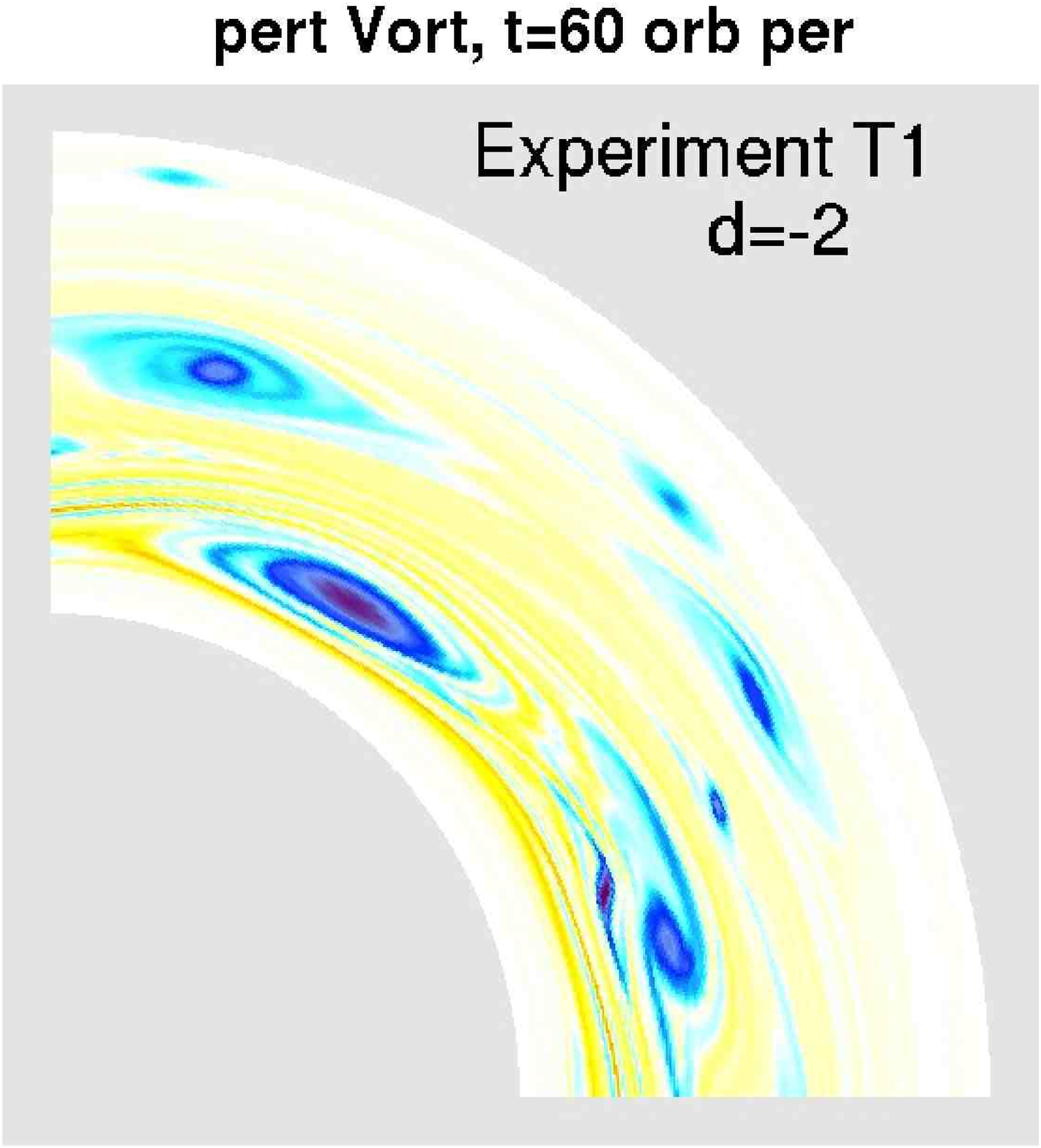}}  
\scalebox{.35}{\includegraphics{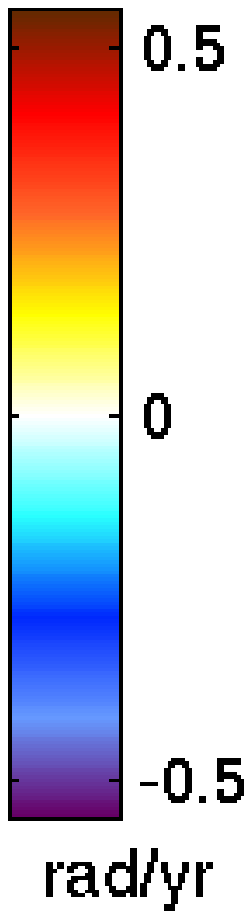}}  
\scalebox{.1}{\includegraphics{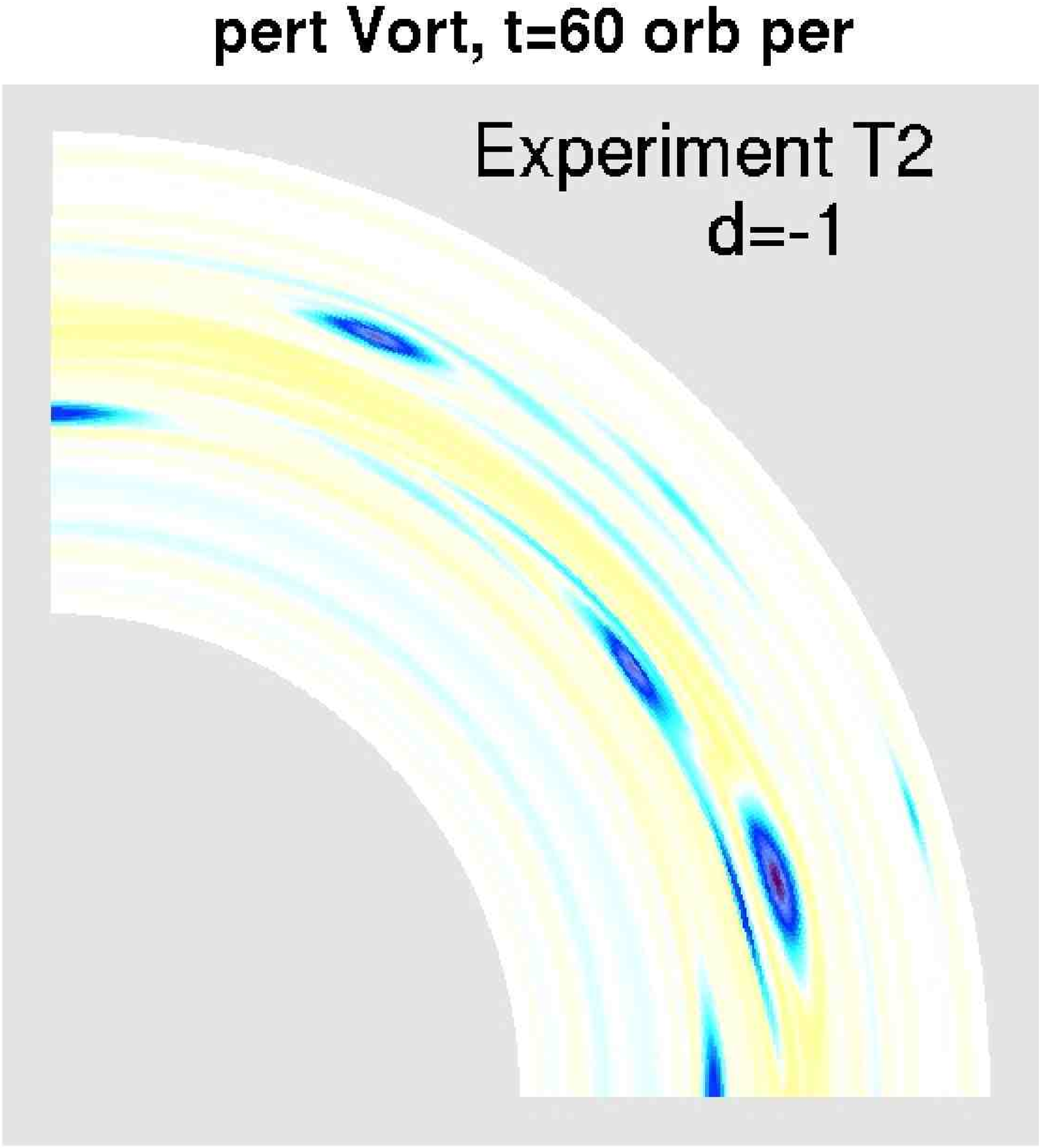}}  
\scalebox{.35}{\includegraphics{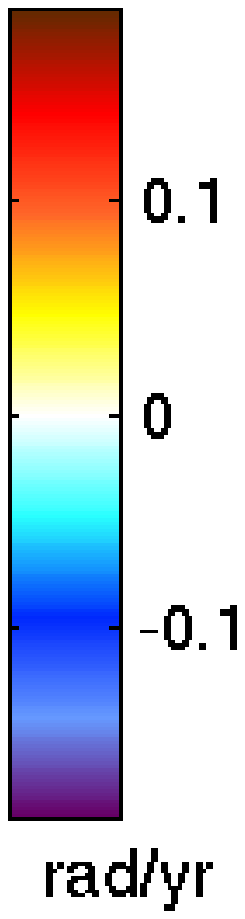}}  
\scalebox{.1}{\includegraphics{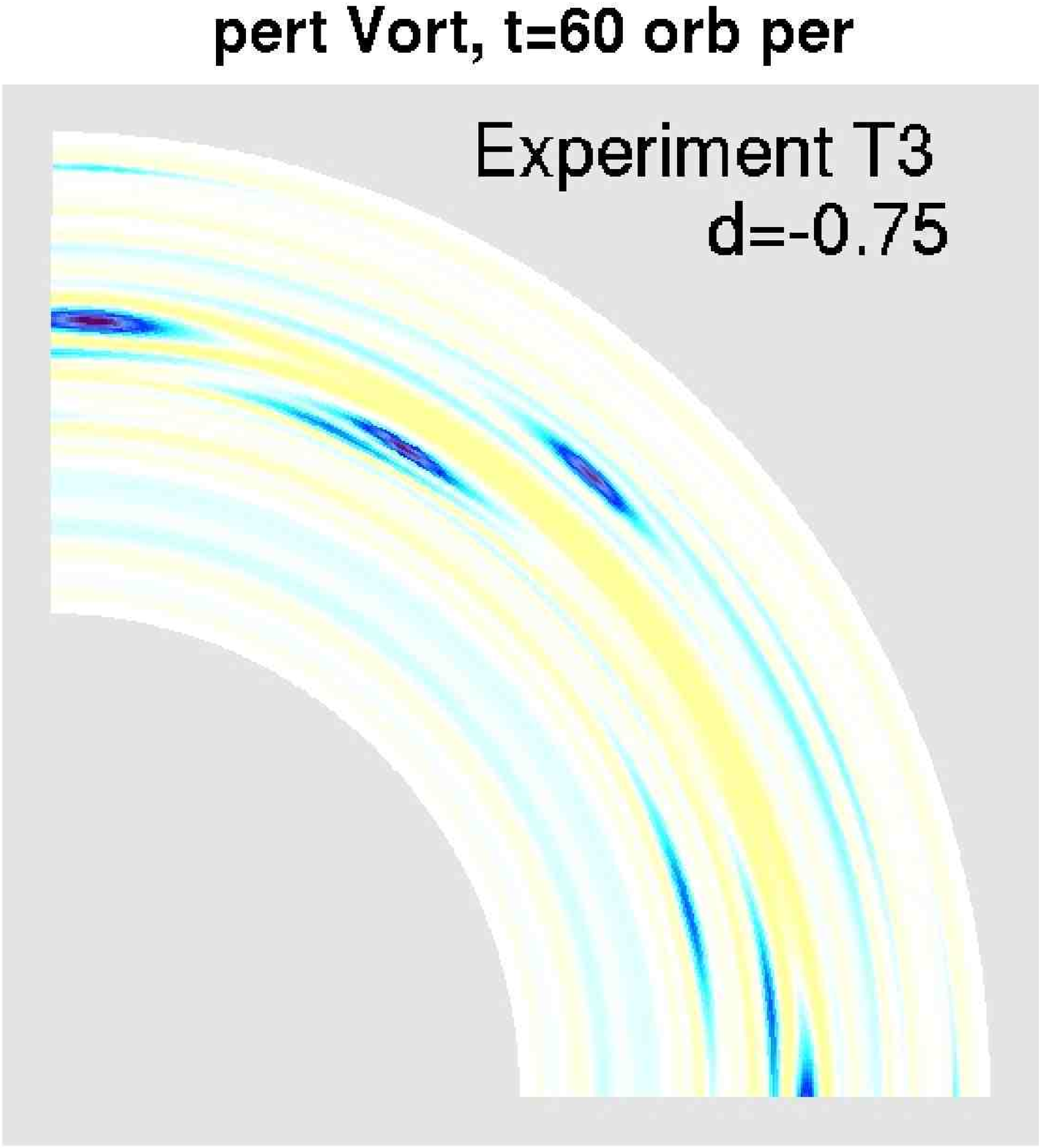}}  
\scalebox{.35}{\includegraphics{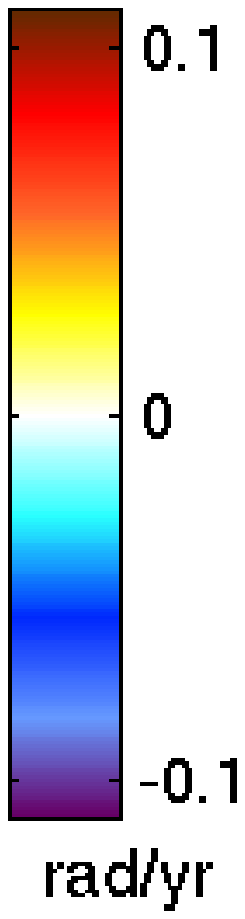}}  
\caption{\label{f_T_sections} 
Perturbation vorticity for simulations T1, T2, and T3, where the background temperature $T_0\sim r\,^d$, and $d=-2$, $-1$, and $-0.75$.  Larger background temperature gradients produce stronger baroclinic instabilities, so that vortices grow faster.}
\end{figure}

\begin{figure}[tbh]
\center
\scalebox{.8}{\includegraphics{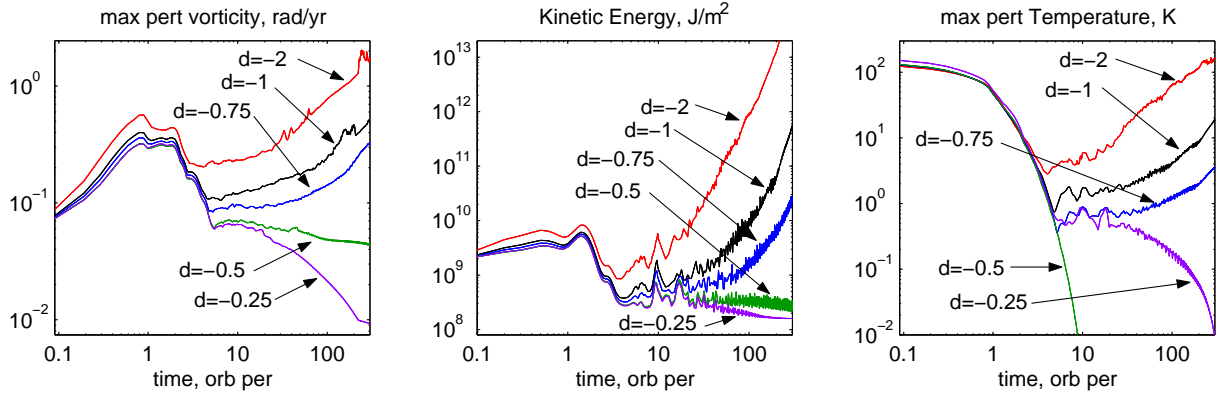}}  
\caption{\label{f_vary_T}  Comparison of data for simulations T1-T5, where the background temperature $T_0\sim r\,^d$, and $d$ ranges from -0.25 to -2.  Simulations where the background temperature gradient is larger in magnitude increases the strength of the baroclinic feedback, resulting in increases in all three measures.}
\end{figure}

\begin{figure}[tbh]
\center
\scalebox{.8}{\includegraphics{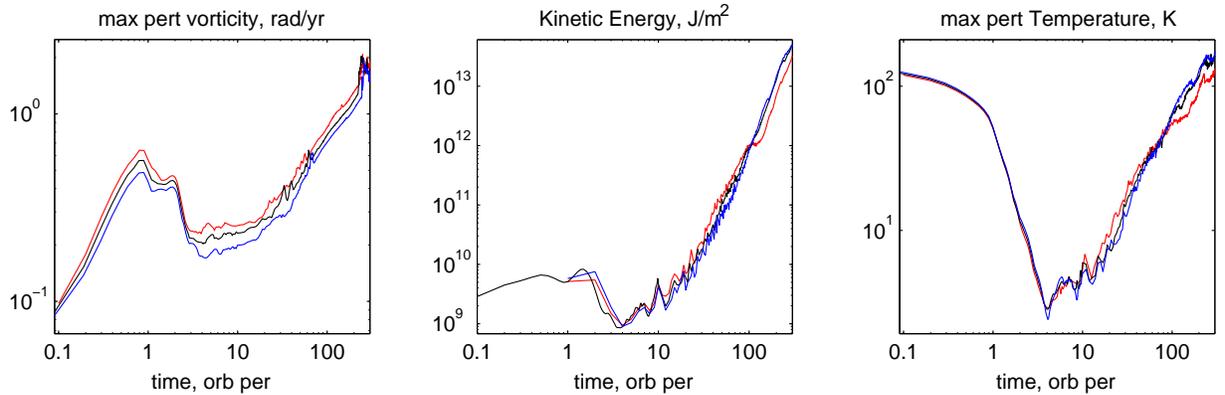}}  
\caption{\label{f_vary_Sigma} Comparison of data for simulations D1-D3, where the background surface density $\Sigma_0\sim r\,^b$ and $b=-1,-3/2$, and $-2$.  This data shows that varying the gradient of surface density results in little difference in the strength of the baroclinic feedback.}
\end{figure}

Other simulations explore the role of background temperature (T1-T5) and background surface density (D1-D3).  
Larger background temperature gradients in simulations T1-T5 result in larger and stronger vortices (Fig. \ref{f_T_sections}).  Quantitative measures such as the kinetic energy, maximum vorticity, and maximum temperature all grow faster with larger temperature gradients (Fig. \ref{f_vary_T}).  The evolution of these quantities does not change as the background density gradient is varied (simulations D1-D3, Fig \ref{f_vary_Sigma}).  It is somewhat surprising that the baroclinic feedback responds strongly to the background temperature gradient but not the background density gradient, when these gradients seem to be on equal footing in the baroclinic term (see eqn. \ref{pres_temp_density}). The background temperature gradient is a source of available potential energy that can be transformed into the kinetic energy of vortex motion as the vortices transport heat from the hot inner disk to the cold outer disk.  This nonlinear heat advection cannot be captured in a linear stability analysis.  Since the surface density is time-independent in our anelastic model, the background surface density gradient cannot provide a source of potential energy for vortex formation.


\subsection{Richardson number}

Several previous studies have used the Richardson number to characterize instabilities in protoplaneary disks \citep{Johnson_Gammie05apj,Johnson_Gammie06apj}, and we compute the Richardson number here for comparison.  We believe that the Solberg-H{\o}iland criterion (\citet{Tassoul00bk,Ruediger_ea02aa}, also see \S 4 of Part I), which was specifically created for differentially rotating astrophysical fluids, is the best way to judge whether a disk is convectively unstable.  For the simulations presented in this paper, the Solberg-H{\o}iland values are positive (0.035--0.299 years$^{-2}$), indicating that they are all convectively stable.  However, the Richardson number also provides useful information about the instability.  We found that the baroclinic feedback is stronger (i.e. vortex growth is faster) in simulations with more negative Richardson numbers.

The Richardson number is often evoked in geophysical turbulence to quantify the relationship between stratification and shear.  For the atmosphere this dimensionless ratio is typically 
\bea
Ri(z) = \frac{N^2(z)}{\left(\dd{u}{z} \right)^2}
= \frac{\ds-\frac{g}{\rho}\frac{d\rho}{dz}}{\left(\dd{u}{z} \right)^2}
\eea
where $N(z)$ is the local Brunt-V\"ais\"a\"a buoyancy frequency, $u$ is the horizontal velocity, $z$ is the vertical coordinate, $\rho$ is the density, and $g$ is the gravitational force \citep{Turner73bk}.  The numerator $N^2$ gives the strength of stratification, where $N^2$ is negative for a convectively unstable fluid, is positive and small for weakly stable stratification, and is positive and large for strongly stable stratification.  The denominator gives the strength of the shear. 

\begin{figure}[tbh]
\center
\scalebox{.8}{\includegraphics{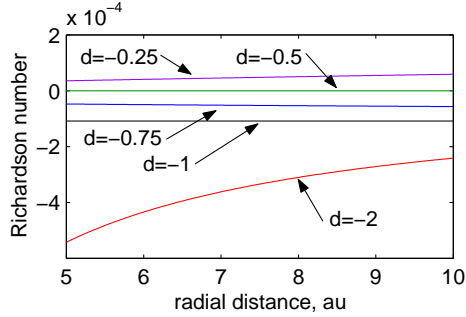}}  
\caption{\label{f_Ri} Richardson number, $Ri(r)$, for simulations T1 through T5, where the background temperature $T_0\sim r\,^d$.  The Richardson number depends only on background functions, and so is constant in time.  By comparing to Fig. \ref{f_vary_T}, one can see that a more negative $Ri$ (exp. T1, $d=-2$) indicates a stronger instability, and less negative $Ri$ (exp. T3, $d=-0.75$) indicates weaker instability.  For exp. T4 ($d=-0.5$) and T5 ($d=0.25$) $Ri$ is zero and positive, respectively, and the baroclinic instability is not active in both cases. }
\end{figure}

In our equation set the Richardson number is
\bea
Ri = \frac{N^2}{\left(r\dd{\Omega_0}{r} \right)^2}
   = \frac{-c_p \ds\frac{d\theta_0}{dr}\frac{d\pi_0}{dr}}
          {\left(r\dd{\Omega_0}{r} \right)^2}.
\eea
By comparing the Richardson number (Fig. \ref{f_Ri}) with kinetic energy or maximum vorticity (Fig. \ref{f_vary_T}) for simulations T1--T5, it is clear that the Richardson number is an excellent way to predict the strength of the baroclinic feedback.  When $Ri\ge0$ (T4 and T5, where $d=-0.5$ and $-0.25$), kinetic energy and vorticity simply decay away.  When $Ri<0$ (T1--T3, where $d=-2$--$-0.75$), the baroclinic feedback operates and kinetic energy and vorticity grow.  In fact, the simulation with the most negative Richardson number (T1, where $d=-2$) also has the fastest vortex growth.


\citet{Johnson_Gammie06apj} found that disks with a nearly-Keplerian rotation profile and radial gradients on the order of the disk radius have $Ri\ge-0.01$, and are stable to local nonaxisymmetric disturbances.  Our simulations are not restricted to this $Ri\ge-0.01$ criterion, as simulations T1-T3 have quickly-growing instabilities but have Richardson numbers in the range of $-5\times10^{-5}$ to $-5\times10^{-4}$.  
The most likely difference between the two models that accounts for this disagreement is that our simulation allows small initial temperature perturbations to evolve into strong local vorticity perturbations that can produce stable vortices.  This initial evolution can only occur if the viscous dissipation is sufficiently low (high Reynolds number).

\subsection{Alpha viscosity}

Protoplanetary disks are often described by the dimensionless number $\alpha$, which is used to parameterize an effective viscosity $\nu = \alpha c_s H_p$ where $H_p$ is the vertical pressure scale height of the disk and $c_s$ is the local sound speed.  This simple description was used to calculate the density structure, temperature structure, and mean components of laminar and turbulent gas flow in a disk (\citealt{Shakura_Sunyaev73aa}, \citealt{Lynden-Bell_Pringle74mnras}, \citealt{Lin_Papaloizou80mnras}).  

Alpha viscosity, rather than Reynolds number, is commonly reported in the astrophysical literature to characterize the dissipation of energy in the disk.  If the pressure scale height is scaled as $H_p=c_s/\Omega_0$, where $\Omega_0$ is the background angular velocity, the alpha viscosity can be calculated as
$\alpha(r) = \nu_e\Omega_0(r)/c_s^2$.
In our anelastic model, this measure cannot be used directly because $c_s>> |{\bf u}'|$ and pressure waves are temporally constrained to adjust instantaneously.  In order to compare the alpha viscosity with other protoplanetary disk models we report the ratio of the alpha viscosity to the Mach number squared,
\bea
\frac{\alpha}{Ma^2} = \frac{\nu_e\;\Omega_0}{{|\bf u}'|^2}
= \frac{\tilde \Omega_0}{Re|{\bf \tilde u}'|^2}
\eea
where $Ma=|{\bf u}'|/c_s$  and tildes indicate nondimensionalized variables.

\begin{figure}[htb]
\center
\begin{tabular}{cc}
\scalebox{.8}{\includegraphics{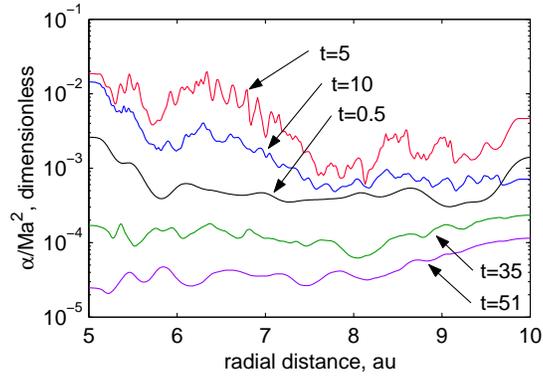}}  
\end{tabular}
\caption{\label{f_alpha_visc} Azimuthal averages of $\alpha/Ma^2$ for a typical simulation (A1) for various times, in orbital periods.  This quotient ranges between $10^{-2}$ and $10^{-5}$ for all simulations.}
\end{figure}

Azimuthal averages of $\alpha/Ma^2$ for all simulations are between $10^{-5}$ and $10^{-2}$ (Fig. \ref{f_alpha_visc}).  Mach numbers of 0.01 or 0.1 would produce corresponding alpha viscosity ranges of $10^{-6}$--$10^{-9}$ and $10^{-4}$--$10^{-7}$.  \citet{Klahr_Bodenheimer03apj} report Mach numbers of 0.03 to 0.3 and $\alpha=10^{-2}$ to $10^{-4}$ for their two-dimensional simulations with radial temperature gradients, which have resolutions of $62^2$ and $128^2$.  \cite{Godon_Livio99apj} report a viscosity parameter $\alpha=10^{-4}$ and $10^{-5}$ for their $128^2$ and $256^2$ simulations, respectively.  In general, higher Reynolds numbers (and thus smaller alpha viscosity) can be achieved with higher resolution.  Our simulations have slightly higher resolution ($256^2$ and $512^2$) than other studies, and the effective Reynolds number is also higher ($\sim10^7$) due to the use of hyperviscosity (see Part I, section 3).  These characteristics contribute to a lower effective alpha viscosity, so that our results include numerous fine, small-scale structures such as layers of filaments swirling around the vortices.

\section{Angular Momentum Transport \label{s_angular_momentum_transport}}
The transport of angular momentum is of critical interest in the study of protoplanetary disks.  The traditional view of disk evolution is that angular momentum is transported outward as mass is transported inward.  In Keplerian motion gas near the star has a faster angular velocity than the gas further out.  Turbulence in the gas creates an effective viscosity, so that faster moving gas in the inner disk will speed up slower gas in the outer disk, and the outer fast gas will tend to slow down the inner gas.  Thus angular momentum is transported outward.  As the inner gas slows down, it is no longer rotationally supported at that orbit, and falls toward the star to gain speed.  Thus mass is transported inward.  Similar arguments can be made for particle collisions, which would enhance this process.

The theory of outward angular momentum transport is based on azimuthally uniform dynamics in a viscous disk.  Turbulence and coherent structures may have radically different effects, and is currently a topic of intense scientific interest.  \citet{Klahr_Bodenheimer03apj} report that, just like in laminar flow, turbulence in baroclinic disks transports angular momentum outward and mass inward, while releasing potential energy.  \citet{Li_ea01apj} used a finite volume model of the compressible Euler equations to model Rossby waves and vortex generation, and found that individual vortices transport angular momentum outward.  \citet{Johnson_Gammie05apjB} also found positive angular momentum flux in their compressible shearing sheet model when they used strong initial vorticity perturbations to trigger vortex formation.

As a simple example, consider locally Cartesian coordinates in the radial and azimuthal direction.  A slanted vortex of this local coordinate system could have the stream function
\bea
\Psi' = \alpha \exp \left( \ds -\left( \left(\frac{\phi}{\phi_0} \right)^2
+ \beta r \phi +     \left(\frac{r}{r_0} \right)^2   \right) \right). 
\label{toy_vortex}
\eea
Each streamline is a rotated ellipse centered at the origin with radial extent $r_0$ and azimuthal extent $\phi_0$.  The angle of the ellipse is only affected by $\beta$.  This vortex is superimposed on some background flow, so the perturbation velocities in locally Cartesian coordinates are $u'=-\p_\phi\Psi'$ and $v'=\p_r\Psi'$.  The angular momentum transport of this vortex is
\bea
\Sigma_0\int_{-\infty}^{\infty} u'v'  \; d\phi
 = - \beta \alpha^2 \phi_0 \Sigma_0\frac{\sqrt{2 \pi}}{4} 
   \exp\left( \ds \frac{1}{2}\left( \frac{r}{r_0} \right)^2 
        \left(-4 + \beta^2 \phi_0^2 r_0^2 \right) \right). \label{analytic_uv}
\eea
The sign of this quantity depends only on $\beta$, the angle of the vortex.  Positive and negative vortices with the same orientation have the same angular momentum transport, as the sign of the vortex only affects $\alpha$, a squared quantity in (\ref{analytic_uv}).  This indicates that it is only the orientation of the vortex within the flow that affects whether momentum travels towards the inside or outside of the disk; the direction of rotation of the vortex is inconsequential.  

\begin{figure}[htbp]
\center
\begin{tabular}{ccc}
\scalebox{.6}{\includegraphics{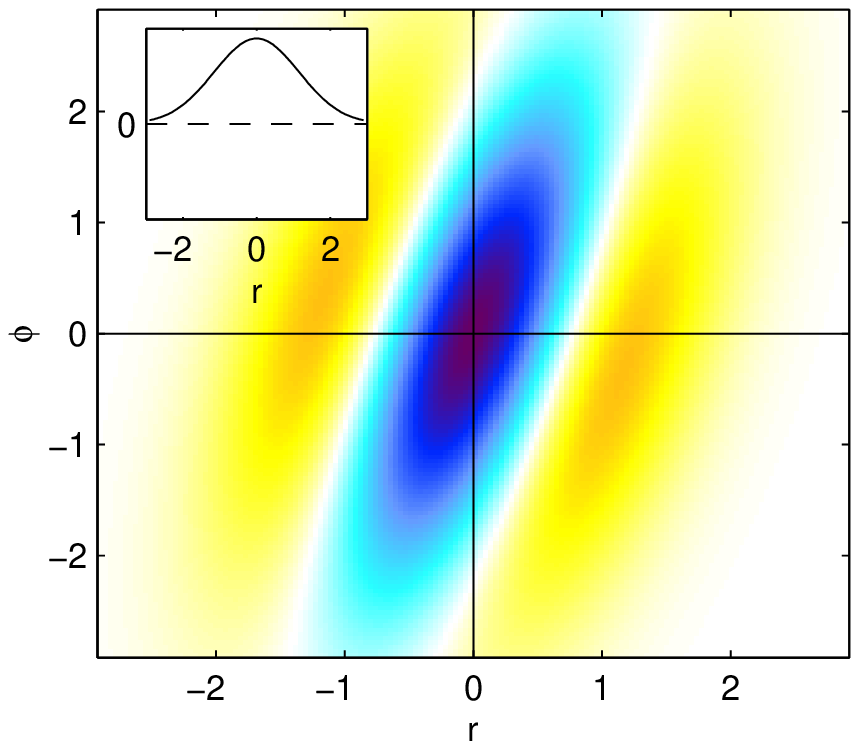}}  
\scalebox{.6}{\includegraphics{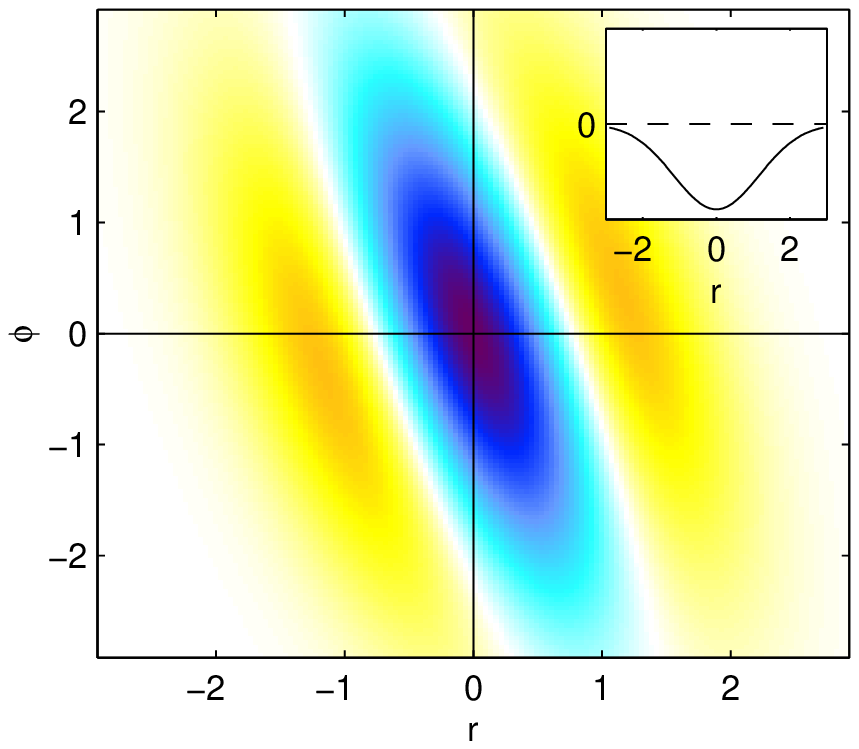}}  
\scalebox{.7}{\includegraphics{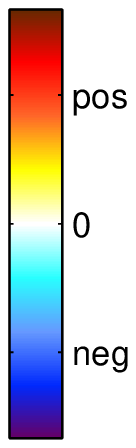}}\\  
\scalebox{.35}{\includegraphics{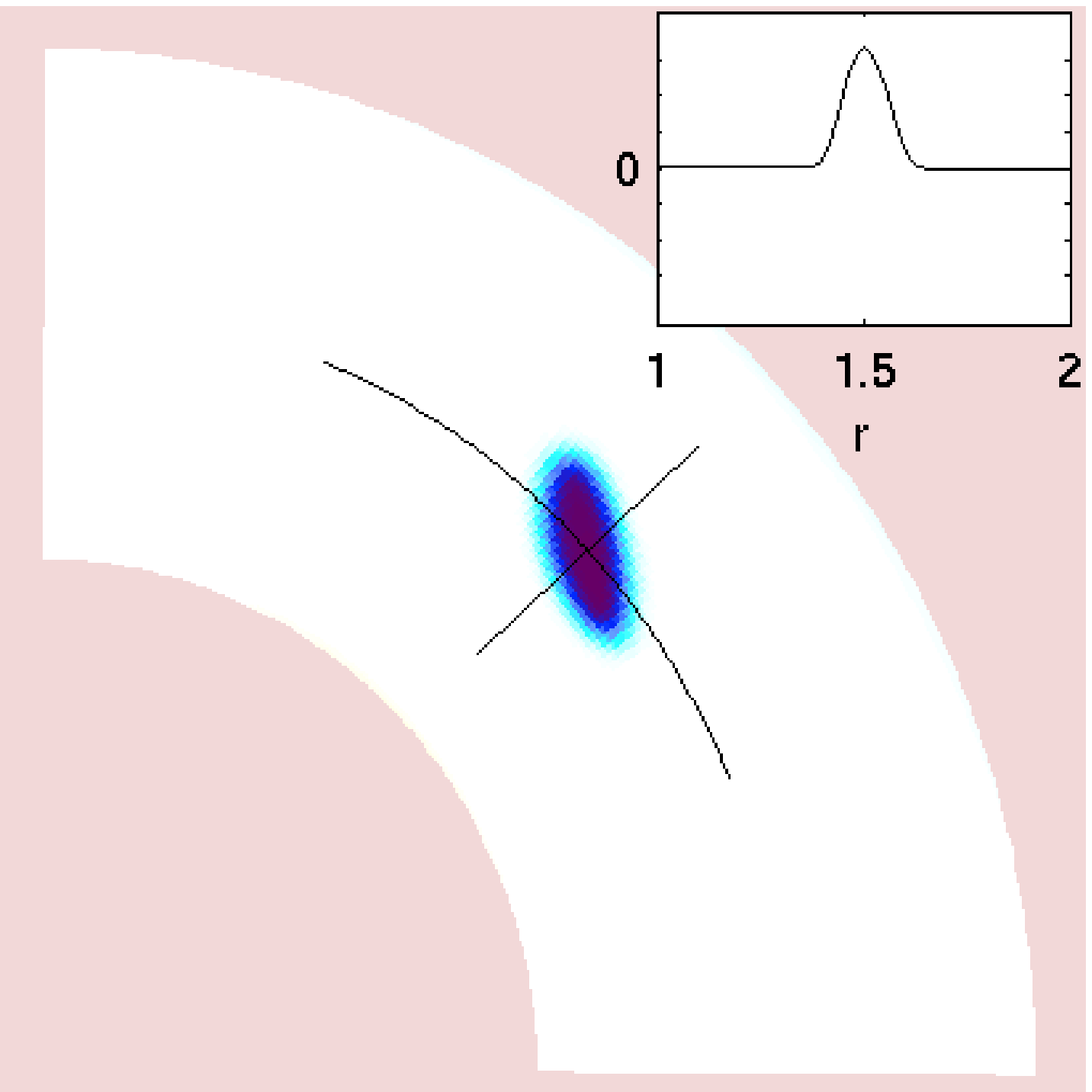}}  
\scalebox{.35}{\includegraphics{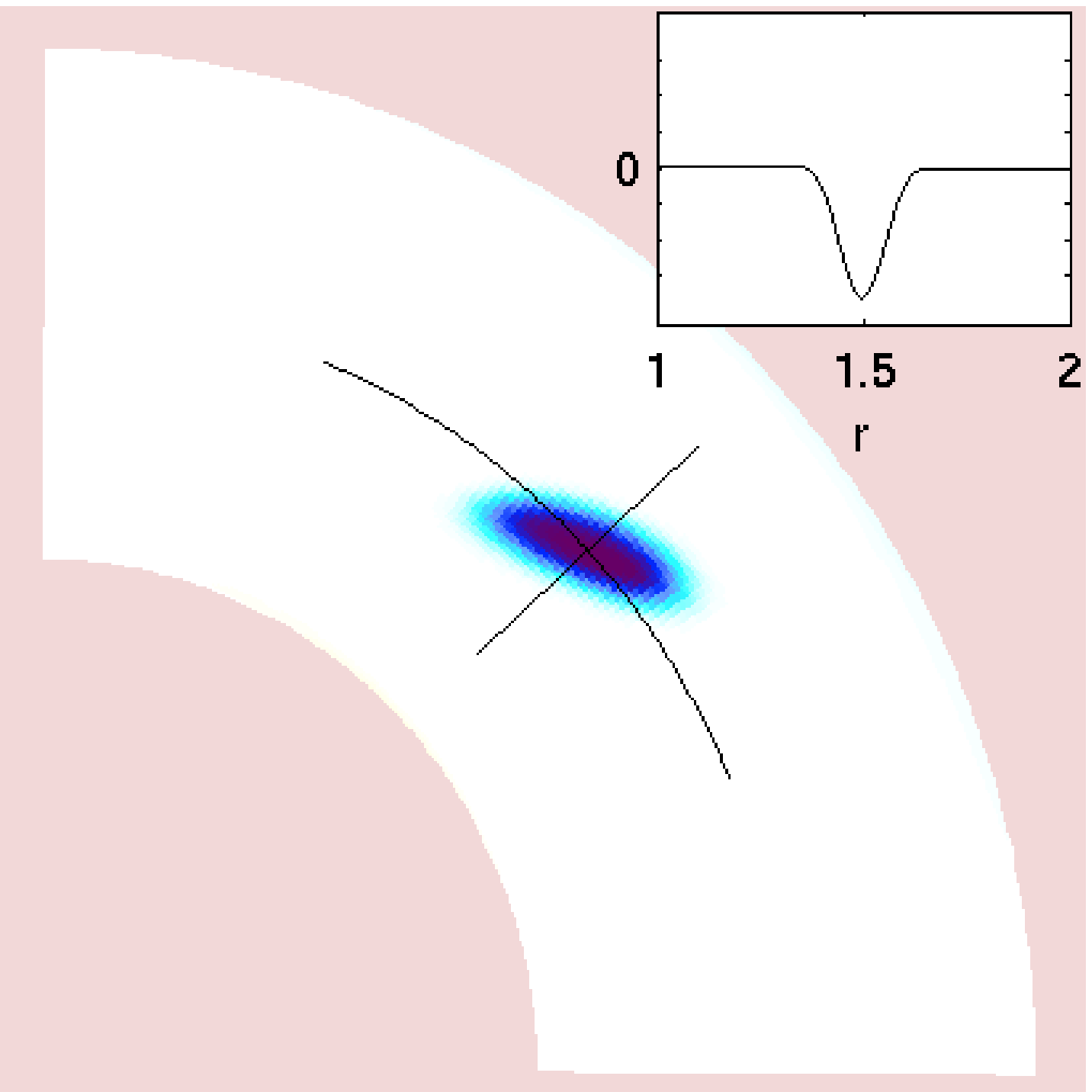}}  
\scalebox{.7}{\includegraphics{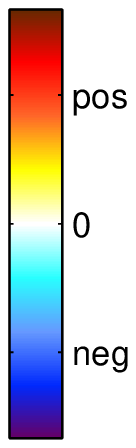}}  
\end{tabular}
\caption{\label{f_ang_mom_ex} Vorticity from the analytic example (top) and numerical model (bottom) where vortices are radially leaning out (left) or leaning in (right).  The inset shows the angular momentum flux, which is positive for outward leaning vortices and negative for inward leaning vortices.  For these pedagogical examples, the scale for the vorticity and angular momentum flux is arbitrary.}
\end{figure}

Our simple analytic example is shown in Fig. \ref{f_ang_mom_ex} for $\beta=-0.5$ (top left) and $\beta=0.5$ (top right), where the other constants are $\phi_0=1$, $r_0=2$,  and $\alpha=1$.  The bottom panels show vortices with similar orientations in the full numerical model.  Clearly, the direction of the angular momentum transport only depends on the angle of the vortex, as in the analytic example.  These vortices are not from the simulations in Table \ref{t_parameters}, but are from short simulations that were specifically designed to produce these orientations.

\begin{figure}[htbp]
\center
\begin{tabular}{cc}
\scalebox{.8}{\includegraphics{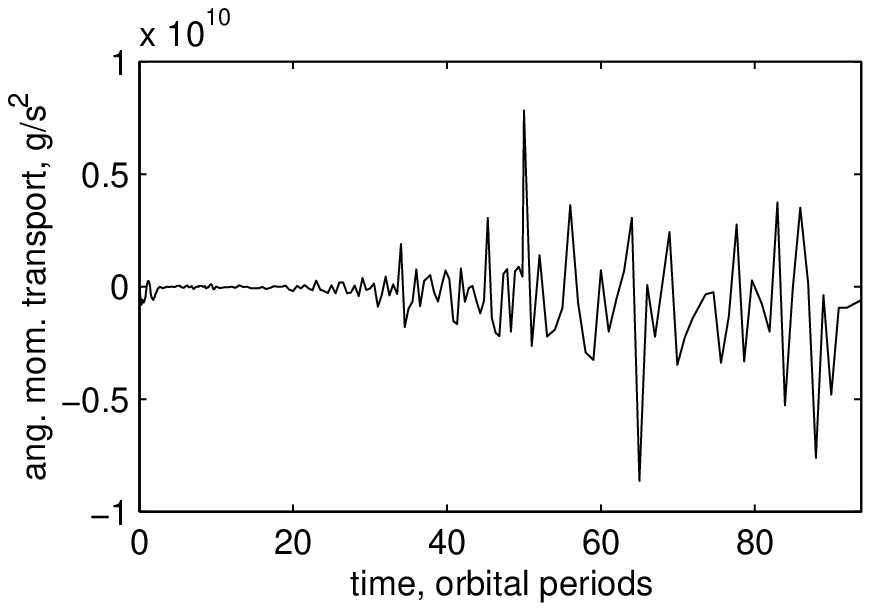}}  
\scalebox{.8}{\includegraphics{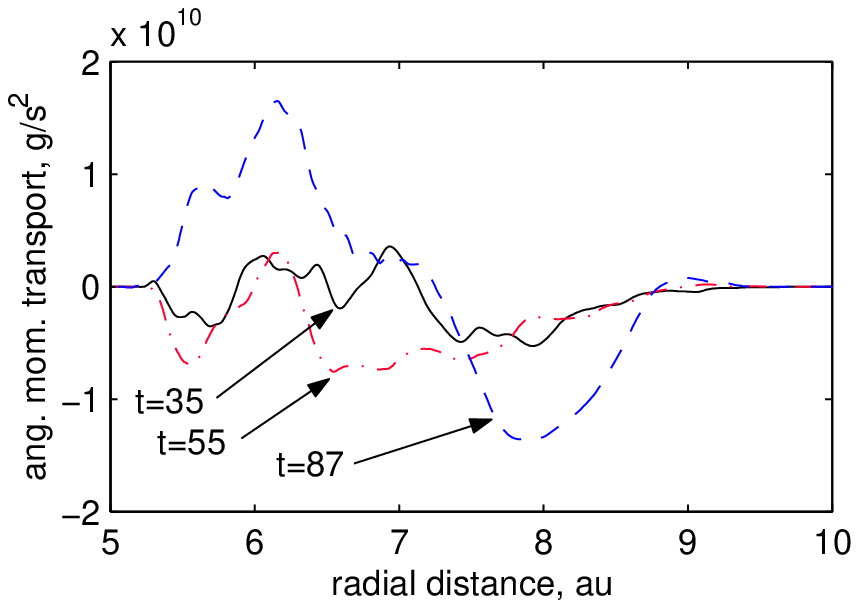}}  
\end{tabular}
\caption{\label{f_ppd_ang_mom} Global angular momentum transport (left) and angular momentum transport as a function of radius (right) for simulation A1, where each curve is an average of ten measurements taken over one-half orbital period, and times correspond to the images shown in Fig. \ref{f_t100}.  As described in the text, angular momentum transport due to a particular vortex depends on the orientation of the vortices.  The spatial and temporal variability in angular momentum transport shown here is due to variability in the orientation of the vortices. }
\end{figure}

What is the effect of vortices on angular momentum transport when they are imbedded in a turbulent flow populated with filaments and other interacting vortices?  To investigate this, the angular momentum transport, $\Sigma_0 u'v'$, was recorded using azimuthal and global averages in the numerical model.  In typical simulations, like A1, the angular momentum transport is highly variable in space and time (Fig. \ref{f_ppd_ang_mom}).  Specifically, the global angular momentum transport cycles chaotically between positive and negative periods as the vorticity field evolves.   The angular momentum transport in these simulations is influenced by the interaction of numerous vortices and vorticity filaments, which is much more complicated than the single vortex case.  We would expect that individual vortices within this flow would contribute angular momentum based on their orientation, and that these individual contributions could be summed to find the angular momentum transport.  However, a separate study of a small number of vortices and filaments in the flow would be required to say conclusively.

Based on the simulation results in Fig. \ref{f_ppd_ang_mom}, we conclude that the total angular momentum transport in an {\it anelastic-gas} turbulent disk with vortices and vortex filaments may be inward or outward, and can vary locally in the disk depending on the orientation of the vortices.  In contrast, studies of {\it compressible-gas} disks have all found that vortices transport angular momentum outwards \citep{Klahr_Bodenheimer03apj,Li_ea01apj,Johnson_Gammie05apjB}.  Compressible-gas models include acoustic waves, which are filtered out of our anelastic model.  Shocks produced by acoustic waves in these studies may orient the vortices uniformly so that angular momentum is transported outwards, or transport angular momentum by other means.

\section{Conclusions \label{s_conclusions}}

In this study we are interested in exploring the necessary conditions for vortex formation in an anelastic protoplanetary disk model that includes baroclinicity and radiative cooling.  We have shown that long-lived vortices can be formed by initial random temperature perturbations through the mechanism of the baroclinic instability.  Vortex production must compete with the strong inhibiting effects of Keplerian shear, an effect observed by other authors (\citealt{Bracco_ea99pf}, \citealt{Godon_Livio99apj}).  Only anticyclones survive in Keplerian disks, while cyclones shear out and diffuse away.  
Many previous studies do not include baroclinic effects due to a lack of thermodynamics \citep{Bracco_ea99pf,Umurhan_Regev04aa, Johnson_Gammie05apjB} or an assumed polytropic relation \citep{Godon_Livio99apj}, and therefore only model decaying turbulence from an initial vorticity distribution.

In the baroclinic feedback, local azimuthal temperature gradients produce vorticity through the baroclinic term in the vorticity equation.  This strengthens vortices, which advect the surrounding thermally stratified gas, producing stronger local temperature gradients.  

The baroclinic feedback can only operate once vortices have formed, as a coherent vortex is required to produce the ``sandwich'' pattern of warm and cold gas about each vortex.  In our simulations two distinct stages can be seen: vortex formation, where the initial temperature perturbation rapidly decays; and vortex growth, where the baroclinic feedback takes effect and both perturbation vorticity and perturbation temperature grow for the rest of the simulation.

The conditions required for the baroclinic feedback are: (1) a sufficiently large radial temperature gradient in the background stratification; and either (2) a fast radiative cooling time; or (3) high thermal dissipation (i.e. small Peclet number).  If the background radial temperature gradient (condition 1) is too small, advection by the vortices does not strengthen local azimuthal temperature gradients.  Both conditions 2 and 3 allow temperature perturbations to track vortices so that the structure of the vorticity and temperature fields are strongly coupled.  The difference between the two mechanisms is that thermal dissipation smoothes out small scale features, resulting in large-scale thermal perturbations, while radiative cooling affects all scales equally and produces smaller thermal perturbations.  Varying the background surface density gradient had no effect on the strength of the baroclinic feedback.

One of the goals of this study was to see if the baroclinic instabilities found by \citet{Klahr_Bodenheimer03apj} can be reproduced in an anelastic equation set with simplified dynamics.  They found that if the background radial entropy gradient is zero---this turns off all baroclinic effects---then the initial vorticity perturbation just decays away (their Model 2).  When entropy varies radially so that the temperature $T\sim r^{-1}$, the flow becomes turbulent within a few orbits and vortices are formed (their Models 3--6).  Our results (condition 1, above) agree with this result, and furthermore show that vorticity grows faster with steeper background radial temperature profiles.

Our conditions 2 and 3 state that thermal dissipation is required for the baroclinic instability.  Indeed, in our simulations where both forms of thermal dissipation were sufficiently slow, the vorticity dies off after the initial vortex formation.  This requirement is in disagreement with the findings of \citet{Klahr_Bodenheimer03apj}, as they ``got rid of radiation transport'' (i.e., there is no radiative cooling) for their 2D simulations.  Because we use a simplified one-parameter radiative cooling model, we can see that the baroclinic feedback strongly depends on the cooling time $\tau$.  It is not clear to us why \citet{Klahr_Bodenheimer03apj} see vortex growth when radiative cooling is missing.  Either there is some implicit thermal diffusion in their code, or their vortices are growing through a different mechanism than ours.

The range of Richardson numbers at which we form vortices differs from \citet{Johnson_Gammie06apj}, who find that simulations with $Ri\ge-0.01$ are stable to local nonaxisymmetric disturbances.  Our simulations with $-0.01\le Ri<0$ form turbulent instabilities and vortices quite easily.  A likely explanation for this difference is that our simulations have the large Reynolds numbers required to permit small initial temperature perturbations to evolve into strong local vorticity perturbations before they are viscously damped.   

The results of a model must be understood within the asymptotic regime where it is valid.  Our model assumes that the disk is thin and hydrostatically balanced in the vertical, so that only large-scale horizontal motions are considered.  The vertical dynamics, which we don't consider, can affect vortex stability as well: \citet{Knobloch_Spruit86aa} argue that height variations must be included when discussing shear instabilities in the disk, because vertical gradients of azimuthal velocity are not small; \citet{Barranco_Marcus05apj} found that columnar vortices are unstable to small perturbations, but that internal gravity waves naturally create robust off-midplane vortices (see discussion in conclusions of Part I).  
%
Indeed, the most serious limitation of this study is our assumption of a 
two-dimensional disk.  If the initial baroclinic instabilities have 
radial scales that are small compared to the disk scale height, as is 
suggested by our local simulations presented in paper I, then the 
vertical stratification of the disk will likely play a major role in 
the nonlinear development and the longevity of vortices.  Nevertheless, 
we believe that our two-dimensional simulations have served to identify 
physical processes that will likely play a role in a fully 
three-dimensional simulation.  For example, if vortices primarily form 
on the upper and lover surfaces of disks, then the radiative cooling 
rates will likely be more rapid than if the vortices were buried in the 
optically thick midplane of the disk.  We therefore expect that 
vortices confined to the surface of a disk could have longer lifetimes 
due to their ability to efficiently transport heat radially outwards.

Our model is based on an anelastic mass conservation equation, which filters out acoustic waves.  Thus shock waves and their potential interactions with vortices and angular momentum transport do not appear in our study, as they have in compressible gas models \citep{Li_ea01apj, Klahr_Bodenheimer03apj, Johnson_Gammie05apjB}.  These studies all found that angular momentum is always transported outward, while we found that it may be inward or outward, and is highly variable in space and time.  The obvious difference in the dynamics is the lack of shocks in our anelastic-gas model.  This suggests that shocks play an important role in the transport of angular momentum in protoplanetary disks.

The baroclinic feedback enhances vortices so that they can be long-lived.  In our simulations they survived for the duration of the longest numerical simulation---for over 600 orbital periods (12,400 years)---and show no signs of decaying.  This study shows that the baroclinic feedback is a viable mechanism for the generation and persistence of vortices in protoplanetary disks.  
In the baroclinic feedback, the background temperature gradient provides a source of available potential energy that can drive the vortices indefinitely even in the presence of a finite rate of viscous dissipation.  As vortices are efficient at collecting particles from the surrounding gas, they are a natural place for planets to form in the disk.  If vortices are long-lived coherent structures in protoplanetary disks, as suggested by this work, they offer a way to overcome the difficulties presented by current planetary formation theories: the high particle concentrations in vortices speed up the core accretion process; likewise, this high particle concentration could lead to gravitational instability.  Both core accretion and gravitational instability are hindered in the majority of the disk where particle concentrations are low.
Strong concentration of particles may require the vortices 
to grow large compared to the scale height of the disk so that they can 
extend through the midplane of the disk where most particles will 
reside.  Fully three-dimensional simulations are therefore required to 
establish of relevance of vortices to planet formation.

\section{Acknowledgements}
We thank P. Marcus for insightful feedback and practical advice,
A.P. Boss for useful discussions, and an anonymous referee for criticism which significantly improved the final version.
MRP has been supported by an NSF Vigre Grant,
DMS-9810751, awarded to the Applied Mathematics Department at the
University of Colorado at Boulder.  KJ has been supported by NSF grant
OCE-0137347 as well as the University of Colorado Faculty Fellowship. 
GRS was supported by NASA's Origins of Solar Systems research program.


\newpage
\bibliographystyle{apj}
\bibliography{planetary_disk,cfd,qg,dynamics,updf_qg,my_pubs}

\clearpage

\newpage

\clearpage

 \begin{table}[btp]
 \center
  \begin{tabular}[c]{c|cccccccc}
   name & grid & $\tau$ & $d$ & $b$ 
        & $Re$ & $Pe$ & endtime \\
   \hline
   A1 & ${\bf 512^2}$   &  1 & {\bf -2} &  -3/2  
      & 4e7 & 4e7 & 100 \\
   A2 & ${\bf 512^2}$   &  1 & {\bf -1} &  -3/2  
      & 4e7 & 4e7 & 60 \\
   A3 & ${\bf 512^2}$   &  1 & {\bf -0.75} &  -3/2  
      & 4e7 & 4e7 & 60 \\
   \hline
   B1 & $256^2$   &  1 & -0.75&  -3/2 
      & 2e7 & 2e7  &300  \\
   \hline
   Tau1 & $256^2$   & {\bf 1} & -2 &  -3/2  
      & 2e7 & 2e7  & 300 \\
   Tau2 & $256^2$   & {\bf 3} & -2 &  -3/2  
      & 2e7 & 2e7  & 300 \\
   Tau3 & $256^2$   & {\bf 10} & -2 &  -3/2  
      & 2e7 & 2e7  & 300  \\
   Tau4 & $256^2$   & {\bf 100} & -2 &  -3/2  
      & 2e7 & 2e7  & 300  \\
   \hline
   T1 & $256^2$   &  1 & {\bf -2}&  -3/2 
      & 2e7 & 2e7  & 300 \\
   T2 & $256^2$   &  1 & {\bf -1}&  -3/2 
      & 2e7 & 2e7  & 300  \\
   T3 & $256^2$   &  1 & {\bf -0.75}&  -3/2 
      & 2e7 & 2e7  &{\bf 600}  \\
   T4 & $256^2$   &  1 & {\bf -0.5}&  -3/2 
      & 2e7 & 2e7  & 300  \\
   T5 & $256^2$   &  1 & {\bf -0.25} &  -3/2  
      & 2e7 & 2e7  & 300  \\
   \hline
   D1 & $256^2$   &  1 & -2 &  {\bf -1}  
      & 2e7 & 2e7  & 300  \\
   D2 & $256^2$   &  1 & -2 &  {\bf -3/2}  
      & 2e7 & 2e7  & 300\\
   D3 & $256^2$   &  1 & -2 &  {\bf -2} 
      & 2e7 & 2e7  & 300  \\
   \hline
   Pe1 & $256^2$   &  100 & -2 &  -3/2 
      &  2e7& {\bf 2e7 }  & 300 \\
   Pe2 & $256^2$   &  100 & -2 &  -3/2 
      & 2e7 & {\bf 1e6}  & 300  \\
   Pe3 & $256^2$   &  100 & -2 &  -3/2 
      & 2e7 & {\bf 1e4}  & 300  \\
 \end{tabular}
 \caption{\label{t_parameters} Model parameters for the numerical simulations discussed in this paper.  Here $\tau$ is the radiative cooling time in orbital periods, and $d$ and $b$ are the powers on the background temperature and surface density functions, $Re$ and $Pe$ are the Reynolds and Peclet numbers, and endtime is in orbital periods.  In simulation B1,  the baroclinic term is turned off at various times during the simulation.
} \end{table}

\end{document}